\documentclass[twocolumn,superscriptaddress,nofootinbib,amsmath,amssymb,aps,floatfix]{revtex4-2}
\usepackage{graphicx,tikz} 
\usepackage[mode=buildmissing]{standalone} 
\usepackage{dcolumn}
\usepackage{bm}
\usepackage{physics}
\usepackage{xcolor}
\usepackage{siunitx}
\usepackage{soul}
\usepackage{comment}
\usepackage[bb=boondox]{mathalfa}
\usepackage{lipsum}

\usepackage{hyperref}
\usepackage{orcidlink}

\begin{document}

\title{Dynamics and pseudospectra around quantum exceptional points}
\author{Georgios~E.~Pavlou\,\orcidlink{0000-0003-4478-935X}}
\email[Contact author: ]{gepavlou@iesl.forth.gr}
\affiliation{Institute of Electronic Structure and Laser (IESL), FORTH, 71110 Heraklion, Greece}
\affiliation{ITCP-Department of Physics, University of Crete, 70013 Heraklion, Greece}
\author{Ioannis~Kiorpelidis\,\orcidlink{0009-0000-9775-8958}}
\affiliation{Institute of Electronic Structure and Laser (IESL), FORTH, 71110 Heraklion, Greece}
\affiliation{ITCP-Department of Physics, University of Crete, 70013 Heraklion, Greece}
\author{Georgios~K.~Simopoulos\,\orcidlink{0009-0004-8957-3784}}
\affiliation{Institute of Electronic Structure and Laser (IESL), FORTH, 71110 Heraklion, Greece}
\affiliation{ITCP-Department of Physics, University of Crete, 70013 Heraklion, Greece}
\author{Manas~Kulkarni\,\orcidlink{0000-0003-1216-1840}}
\affiliation{International Centre for Theoretical Sciences, Tata Institute of Fundamental Research, Bangalore 560089, India}
\author{Konstantinos~G.~Makris\,\orcidlink{0000-0002-0626-0589}}
\affiliation{Institute of Electronic Structure and Laser (IESL), FORTH, 71110 Heraklion, Greece}
\affiliation{ITCP-Department of Physics, University of Crete, 70013 Heraklion, Greece}

\begin{abstract}
Exceptional points (EPs) are a hallmark of non-Hermitian physics, giving rise to extreme sensitivity and enhanced dynamic response. Such unique spectral singularities have been extensively studied in classical photonics, both at theoretical and experimental levels, where the relevant observables are evident. That is not necessarily the case in the quantum regime, where this kind of behavior around EPs, to a large extent, remains elusive. In fact, the interplay between dissipation and quantum fluctuations makes the complete Lindbladian evolution quite different from effective non-Hermitian Hamiltonian descriptions. In this context, we investigate the signatures of the Liouvillian EPs on spectral and dynamic response by considering properly normalized long-time averages and imbalances. Even though our approach is generic, we focus on two paradigmatic models—a lossy Bose-Hubbard lattice and a dissipative two-spin system, where we investigate the role of quantum jumps and interactions in the Liouvillian spectrum, in terms of complex and structured pseudospectra. Our results demonstrate how non-Hermitian phenomena emerge in realistic quantum lattices, providing a bridge between classical EP physics and experimentally
accessible open quantum systems.
\end{abstract}


\date{\today}

\maketitle

\section{Introduction} 

Open quantum systems are physical models that are coupled to their surrounding environment, leading to dissipation, decoherence, and irreversible dynamics \cite{Breuer2007}. Proper physical description of such effects is quite different from the unitary evolution of closed systems and they play a central role in a wide range of physical settings, from condensed matter systems and quantum statistical mechanics \cite{Prosen2011,Eisert2015} to quantum information processing \cite{Nielsen2010} and quantum technologies \cite{Lau2018,Couteau2023}. Moreover, techniques developed in the context of open quantum systems have proven powerful in areas such as quantum optics \cite{Haken1981,Carmichael1993}, quantum cosmology \cite{Kiefer2007}, quantum thermodynamics \cite{Dann2021}, quantum many-body systems \cite{Fazio2025}, and other fields \cite{Rotter2015}. The standard theoretical framework for describing these processes is provided by the Lindblad master equation framework, which governs the time evolution of the system’s density matrix by incorporating both coherent dynamics and environmentally induced dissipation \cite{Lindblad1976} if the interaction with the environment is Markovian or memoryless. This formalism has become a cornerstone for understanding realistic quantum dynamics, where environmental coupling is not merely a perturbation and determines the overall behavior of the system.

In this context, effective descriptions of open quantum systems inevitably lead to non-Hermitian Hamiltonians (NHH) \cite{Moiseyev2011}. Such non-Hermitian formulations have a long history, with early developments in parity-time ($PT$) symmetric quantum mechanics demonstrating that non-Hermitian operators can exhibit entirely real spectra under specific conditions, as well as rich and unconventional physical behavior \cite{Bender1998}. { Their use, however, substantially predates the recent activity in $PT$ symmetric systems. For example, complex optical potentials have long provided a successful description of nuclear scattering and reaction channels in the optical model \cite{Dickhoff2019,Atkinson2025}, where, for example, the imaginary part of the potential represents flux lost from the elastic channel into other channels. Moreover, projection-operator methods, such as the Feshbach formalism, yield effective Hamiltonians after eliminating continuum or reaction channels \cite{Feshbach1962}. Finally, complex-coordinate rotation/complex scaling turns atomic and molecular resonances into complex eigenvalue problems \cite{Ho1983,Myo2026}. The original Hamiltonian may be Hermitian, but complex-coordinate rotation produces a non-Hermitian complex-scaled Hamiltonian whose complex eigenvalues describe resonances. At this point, however, we have to emphasize that the recent intense research activity in non-Hermitian physics was actually triggered by the notion of $PT$-symmetric quantum mechanics and $PT$ optics.}

A defining feature of non-Hermitian systems is the existence of exceptional points (EPs), where both eigenvalues and eigenvectors coalesce, leading to nondiagonalizable matrices \cite{Berry2004,Heiss2004a}. These unique degeneracies give rise to nontrivial spectral and dynamical characteristics, and provide a potential physical platform, where novel effects could be realized. In this respect, apart from a few early studies in microwave cavities \cite{Richter2001} and electronic circuits \cite{Heiss2004b}, it was not until recently that such non-Hermitian Hamiltonians were theoretically proposed \cite{Ganainy2007,Makris2008,Musslimani2008,Makris2010}, and experimentally demonstrated \cite{Guo2009,Ruter2010,Ozdemir2019,Xia2021,Xiao2021,Klauck2025}, in the context of classical optics, and led to the frontier of non-Hermitian photonics \cite{Ganainy2018}. These systems have revealed a variety of striking phenomena associated with exceptional points, including enhanced sensitivity to perturbations, unidirectional transport, and mode switching. Such effects have been exploited in a range of applications, including ultrasensitive microresonators \cite{Wiersig2014,Wiersig2016,Khajavikhan2017,Yang2017}, nanolasers \cite{Khajavikhan2014}, and exceptional-point gyroscopes \cite{Khajavikhan2019}. { NHH dynamics has also been realized and probed in genuinely quantum platforms, including postselected superconducting-qubit experiments \cite{Naghiloo2019}, measurements of Yang-Lee zeros and related nonunitary criticality in open quantum systems \cite{Gao2024YangLee}, and non-Hermitian quantum walks \cite{Xiao2017QW}.} The high degree of control in these classical and quantum platforms has established them as ideal test beds for probing the consequences of non-Hermitian degeneracies.

Despite these advances in classical photonic systems, the manifestation of EP physics in genuinely quantum regimes remains significantly more subtle \cite{Minganti2019}. In fact, in quantum systems, dissipation is inherently linked to stochastic processes such as quantum jumps (QJs), and the dynamics is governed by the full Lindblad evolution rather than effective non-Hermitian Hamiltonians alone. As a consequence, many signatures of exceptional points that are well established in classical platforms—such as sharp spectral transitions and enhanced sensitivity—are not directly accessible and can be obscured by the underlying quantum dynamics. While recent studies have begun to explore exceptional points in quantum settings, both theoretically \cite{Minganti2020,Khandelwal2021,Thapliyal2025,Kopciuch2025,Lin2025,Downing2025} and experimentally \cite{Wang2021,Ding2021,Han2023,Abo2024,Zhang2022,Bu2023,Gao2025}, a clear understanding of how their characteristic features emerge under realistic open-system evolution is still under investigation. In particular, it remains unclear whether the characteristic features of non-Hermitian Hamiltonians persist in experimentally relevant observables in the presence of QJs. Our motivation is further supported by recent experimental works  \cite{Wang2021,Ding2021,Han2023} that examine postselected evolution in the absence of QJs, highlighting the importance of understanding how QJs modify EP physics.

In this work, we investigate how QJs influence both spectral sensitivity and dynamical behavior near EPs in open quantum systems. By comparing the full Lindblad evolution with the effective non-Hermitian description, we examine how including or omitting QJs affects the eigenvalue bifurcations and dynamical observables. We focus on two paradigmatic models: a lossy two-site Bose-Hubbard dimer \cite{Jaksch1998,Diehl2008,Graefe2008prl,Graefe2008,Daley2014,Kordas2015,Kordas2018,Moca2023,Zheng2024} and a driven-dissipative two-spin system \cite{Prosen2011,Minganti2019,Znidaric2015,Kulkarni2023,Karanikas2024}. Our approach combines a detailed spectral analysis (exact diagonalization)—including Liouvillian eigenvalues and complex/structured pseudospectra \cite{Trefethen2005,Sato2020b,Makris2021,Komis2022,Kiorpelidis2024,Kiorpelidis2026}—with a study of dynamical signatures such as population imbalance, spin polarization, and normalized long-time averages (LTAs). This enables us to determine how QJs suppress coherent oscillations, smoothen transitions across EPs, and simultaneously reshape the sensitivity landscape encoded in the associated pseudospectra. 

This paper is organized as follows: Section \ref{theoretical} outlines the theoretical framework. Sections \ref{bose_hubbard} and \ref{spin_chain} analyze the Bose-Hubbard dimer and the two-spin system, respectively. Section \ref{conper} presents our conclusions. Technical details, regarding the Lindblad master equation, related analytical results, the quantum-jump method (QJM), pseudospectra theory, and additional comparisons between non-Hermitian Hamiltonians and Lindbladian dynamics, are provided in Appendixes \ref{app:linbland}–\ref{app:qjm}.

\section{Theoretical description of open quantum systems and observables}
\label{theoretical}

Let us start by providing the theoretical framework of our study. The dynamics of open quantum systems can be described using the Lindblad master equation \cite{Breuer2007}:
\begin{equation}
\label{master_equation}
\frac{d\rho(t)}{dt} = -\frac{i}{\hbar} [H, \rho(t)] + \sum_k \left( L_k \rho(t) L_k^\dagger - \frac{1}{2} \{L_k^\dagger L_k, \rho(t)\} \right),
\end{equation}
where $\rho(t)$ is the reduced density matrix, $H$ is the Hamiltonian, and $L_k$ are Lindblad operators modeling dissipation and decoherence. This equation preserves the trace of $\rho(t)$ during the time evolution.  The commutator describes unitary evolution, while the dissipative term has two components:
(1) $L_k \rho L_k^\dagger$, which generates QJs, and
(2) the anticommutator term, which represents continuous loss of coherence.
Technical details of the numerical implementation are summarized in Appendix~\ref{app:linbland}.
For any observable $O$, its expectation value is
\begin{equation}
\label{expectation}
\langle O (t) \rangle = \text{Tr}[O \rho(t)].
\end{equation}

In the present work, the term NHH refers specifically to the no-QJ terms associated with the Lindblad master equation in Eq.~\eqref{master_equation} rather than to the most general class of NHHs used in scattering, resonance, projection, or post-selected quantum settings. A commonly used approximation is to neglect the QJs. This leads to the following effective non-Hermitian Hamiltonian (note that we set $\hbar=1$ for the remainder of the paper):
\begin{equation}
\label{nhh}
H_{\text{eff}} = H - \frac{i\hbar}{2} \sum_k L_k^\dagger L_k.
\end{equation}
{ The
inverse problem is not unique: Starting from a general non-Hermitian
Hamiltonian does not, in general, determine a unique Lindblad master equation
or a unique set of jump operators.} In this case, because the trace of $\rho(t)$ decays under $H_{\text{eff}}$, expectation values require the following normalization:
\begin{equation}
\label{expectationnoQJ}
\langle O(t) \rangle = \frac{\text{Tr}[O \rho(t)]}{\text{Tr}[\rho(t)]}.
\end{equation}
{ In Eq.~\eqref{expectationnoQJ}, we use the standard Hilbert-space metric inherited from the Lindblad problem. The NHH considered here has the same physical Hilbert space as Eq. ~\eqref{master_equation} \cite{Breuer2007,Daley2014}. No additional biorthogonal inner product is introduced for the dynamical observables studied below.}

Although such an approximation has been extensively studied, $H_{\text{eff}}$ cannot capture the full dissipative dynamics in general. In other words, the Liouvillian spectrum provides the complete description of our problem (see Appendix~\ref{app:eff} for comparisons).

An equivalent description is provided by the QJM \cite{Breuer2007,Kordas2015}, which simulates stochastic trajectories of the system's wavefunction. Between jumps the state evolves under $H_{\text{eff}}$, while jumps occur at random times dictated by the Lindblad operators.
Averaging over many trajectories reproduces the solution of the master equation.
QJM is particularly valuable for interpreting experiments that rely on postselection or conditional dynamics, and for understanding the role of individual jump events. It also provides an efficient alternative to direct Liouvillian evolution in larger Hilbert spaces. Throughout this work, we focus on the exact Lindblad master equation approach, but we compare our results to QJM in Appendix \ref{app:qjm}.

In order to quantify long-term dynamical behavior, one relevant way is to consider the LTA of an observable:
\begin{equation}
\langle O(\gamma) \rangle_T
= \lim_{T \to \infty}
\frac{1}{T}\int_0^T dt \frac{\text{Tr}[O \rho(t)]}{\text{Tr}[\rho(t)]},
\end{equation}
The normalization in the integrand ensures that the observable remains well defined in the no-QJ (NHH) description, where probability is not conserved. Such a criterion has been theoretically proposed \cite{Brody2012} and used in quantifying the $PT$ symmetry breaking in a recent experiment in ion traps \cite{Wang2021}.

Regarding the characterization of the Liouvillian or the effective Hamiltonian eigenspectrum, a suitable measure is the associated pseudospectrum. Such a method helps us to characterize the sensitivity of the eigenvalues of a non-Hermitian operator to complex or structured perturbations \cite{Trefethen2005}.
Near EPs, weak perturbations can induce disproportionately large differences in eigenvalues, making the pseudospectra a more informative diagnostic method than the eigenspectrum alone.
By measuring how the pseudospectral radius scales with perturbation strength, one can identify the order of an EP and quantify the degree of non-Hermitian sensitivity.
The numerical investigation of these scaling laws is provided in detail in Appendix~\ref{app:pseudospectrum} and plays a crucial role in distinguishing EPs of different orders.

\section{Non-Hermitian bosonic system: Dissipative Bose-Hubbard model}
\label{bose_hubbard}

The first physical system that we consider is the open Bose-Hubbard model, which is a paradigmatic framework for describing $N$ interacting bosons distributed across $M$ lattice sites, including dissipation. It plays a central role in the study of bosonic lattices, quantum phase transitions, and open-system effects in ultracold atomic gases \cite{Jaksch1998,Graefe2008prl,Graefe2008,Kordas2015,Kordas2018,Bloch2008}. In particular, it provides the standard theoretical description of superfluid–Mott-insulator transitions observed in optical lattices \cite{Bloch2008,Greiner2002} and serves as a versatile quantum simulation platform with applications ranging from quantum transport to quantum information processing \cite{Chanda2024}. Within this broad context, the Bose-Hubbard model provides a physical platform to explore how dissipation and QJs modify bosonic dynamics and influence the emergence of EPs in open quantum systems. { Related work on balanced gain-loss Bose-Hubbard dimers showed that many-particle Lindblad dynamics can depart from mean-field descriptions \cite{Dast2017}. Here, however, we address a distinct comparison, namely, between the no-QJ and the full Lindblad evolution, and combine the resulting Liouvillian spectra and pseudospectra with the study of the dynamics near EPs.}

In particular, the coherent (unitary) dynamics of the corresponding closed system is governed by the Bose-Hubbard Hamiltonian \cite{Kordas2015}:
\begin{align}
\label{BHH}
    {H}_{\text{BH}} &= \sum_{j=1}^M \varepsilon_j {a}_j^\dagger {a}_j - \sum_{j=1}^{M-1} J_j \left( {a}_j^\dagger {a}_{j+1} + {a}_{j+1}^\dagger {a}_j \right) \nonumber \\
    &\quad + \sum_{j=1}^M \frac{U_j}{2} {a}_j^\dagger {a}_j^\dagger {a}_j {a}_j,
\end{align}
where $a_j^\dagger$ and $a_j$ denote bosonic creation and annihilation operators on site $j$, $\varepsilon_j$ are on-site energies, $J_j$ are tunneling amplitudes, and $U_j$ denote local interaction strengths. In the closed system, the total particle number $N$ is conserved, and the Hilbert space dimension is $\dim(\mathcal{H}) = \binom{N+M-1}{M-1}$, reflecting the number of ways to distribute $N$ indistinguishable bosons across $M$ sites.

In an open Bose-Hubbard system, interactions with the environment break particle-number conservation and fundamentally alter the dynamics \cite{Kordas2018}. For instance, particle loss at the leftmost site is modeled by the Lindblad operator
\begin{equation}
\label{loss}
L_1 = \sqrt{\gamma} a_1,
\end{equation}
where $\gamma$ denotes the dissipation rate.

As a minimal but insightful starting point, we focus on a two-site system with a single particle initially prepared on the lossy site. This model is analytically solvable (see Appendix \ref{app:bose}) and thus the competition between tunneling ($J$) and dissipation ($\gamma$), which gives rise to EPs, can be directly investigated. By solving the Lindblad equation, we analyze population dynamics, steady-state behavior, and their dependence on $\gamma$, illustrating how EP physics emerges even in this minimal configuration.

Apart from the noninteracting case, we examine the effect  of interactions with two bosons on two lattice sites, initially localized on the lossy site. This allows us to investigate how interactions modify the EPs, spectral sensitivity, and dynamical evolution, connecting few-body dissipation to broader non-Hermitian phenomena. Unless stated otherwise the parameters we use are $M=2$, $N=1$, $\varepsilon_1=\varepsilon_2=U_1=U_2=0$, and $J=1/2$ for the noninteracting case and the same but with $N=2$ and $U_1=U_2=0.1$ for the interacting one. In the dynamics, we always consider all particles at the lossy site, as our initial condition.

\begin{figure}[h]
\includegraphics[width=1.0\linewidth]{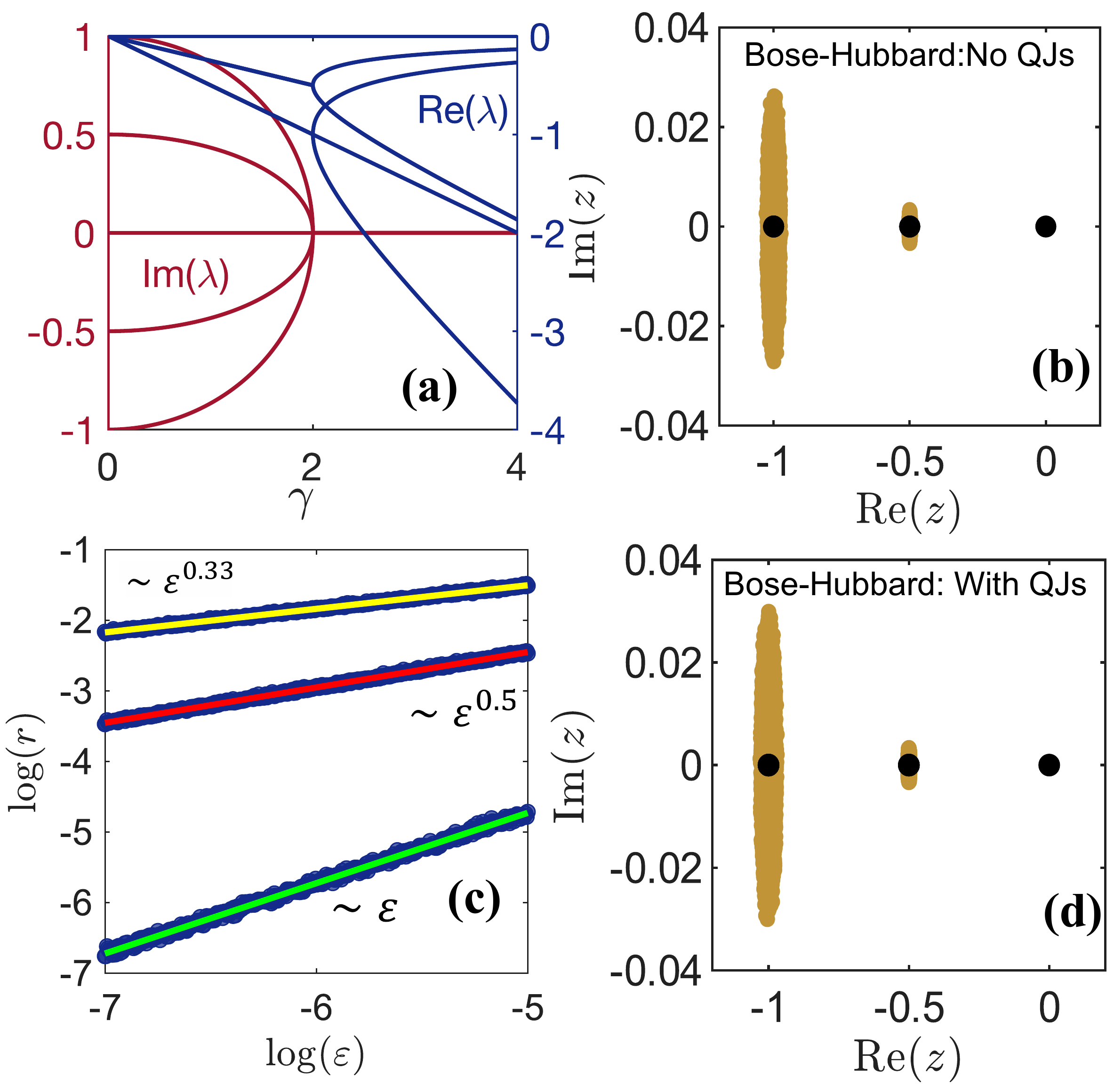}
\caption{Liouvillian eigenspectrum and associated complex pseudospectra of the noninteracting Bose-Hubbard model. 
(a) Eigenvalue spectrum of the Liouvillian superoperator as a function of the dissipation rate $\gamma$. 
The spectrum is identical with and without QJs, and an EP occurs at the analytically known value $\gamma = 4J$. 
(b), (d) Complex pseudospectrum plotted on the complex plane at the EP $\gamma = 4J$ without QJs [panel (b)] and with QJs [panel (d)] ($\epsilon = 10^{-5}$). 
Black dots denote Liouvillian eigenvalues, while yellow dots denote the Liouvillian pseudospectrum, illustrating the enhanced spectral sensitivity near the EP. 
(c) Scaling of the pseudospectral radius $r$ with perturbation strength $\epsilon$ at the EP. 
Only the QJ-included case is shown, as the scaling behavior is identical in both descriptions. 
The extracted power-law exponents correspond to two EPs of second order (red fit, $\sim \epsilon^{1/2}$) and one EP of third order (yellow fit, $\sim \epsilon^{1/3}$), alongside a regular linear scaling (green fit, $\sim \epsilon$). Blue dots in panel (c) denote numerical data. 
}
\label{fig:spectral_analysis_BH}
\end{figure}

\subsection{Liouvillian eigenspectra and complex pseudospectra}

\begin{figure*}[ht]
\includegraphics[width=0.75\linewidth]{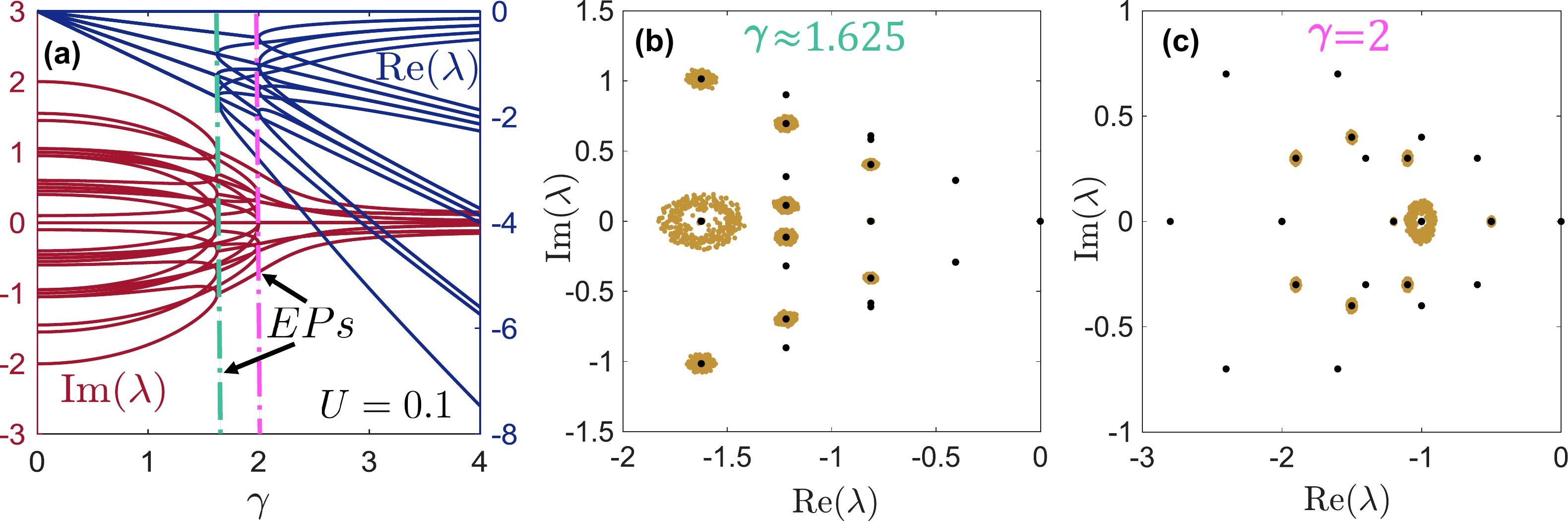}
\caption{Liouvillian eigenspectrum and associated complex pseudospectra of the interacting Bose-Hubbard model. 
(a) Eigenvalue spectrum of the Liouvillian superoperator as a function of the dissipation rate $\gamma$ for the interacting case. 
(b) Complex pseudospectrum at $\gamma \approx 1.625$, revealing enhanced spectral sensitivity associated with eight EP2 degeneracies and one EP3 degeneracy. 
(c) Complex pseudospectrum at $\gamma = 2$, with eight EP2 degeneracies (two of them at $\lambda=-1/2$) and one EP3 degeneracy (both for $\epsilon = 10^{-4}$).}
\label{fig:SpectralU}
\end{figure*}

We begin by analyzing the spectral properties of the Liouvillian superoperator for the noninteracting two-site Bose-Hubbard model with a single particle. The eigenvalue spectrum provides information about decay rates and oscillation frequencies, while the pseudospectrum captures the sensitivity of these spectral features to perturbations—a diagnostic near EPs.

Figure~\ref{fig:spectral_analysis_BH}(a) shows the real and imaginary parts of the Liouvillian eigenvalues as a function of the dissipation rate $\gamma$. Importantly, the spectrum is identical in the descriptions with and without QJs. This holds when the Hamiltonian conserves the number of particles and the Lindblad operators only remove them (no dephasing or gain) \cite{Torres2014}. As $\gamma$ increases, pairs of eigenvalues approach one another and coalesce at the analytically known EP located at $\gamma = 4J$. At this point, both eigenvalues and eigenvectors merge, rendering the Liouvillian nondiagonalizable. Beyond the EP, the spectrum bifurcates into distinct branches, signaling the transition to a qualitatively different dynamical regime. The same EP structure is also present in the effective non-Hermitian Hamiltonian description, as discussed in Appendix~\ref{app:eff}.

The complex pseudospectrum at the EP, shown in Figs.~\ref{fig:spectral_analysis_BH}(b) and \ref{fig:spectral_analysis_BH}(d) for the cases without and with QJs, respectively, is nearly identical in the two descriptions. This indicates that, for the noninteracting Bose-Hubbard model, QJs do not qualitatively modify the spectral sensitivity of the Liouvillian. The eigenvalue bifurcations are therefore fully captured by the Liouvillian spectrum itself.

We now turn to the interacting case, where two bosons occupy the two-site Bose-Hubbard system and interact via a finite on-site interaction, leading to richer eigenvalue bifurcations. Figure~\ref{fig:SpectralU}(a) shows the Liouvillian eigenvalue spectrum as a function of the dissipation rate $\gamma$ for $U=0.1$. In contrast to the noninteracting case, interactions lift certain degeneracies and generate multiple EPs of different orders (higher-order EPs of order $N$ are denoted as EPNs) distributed across the spectrum. While the overall spectral evolution with $\gamma$ remains continuous, eigenvalue coalescences now occur at several distinct values of $\gamma$, reflecting the interplay between tunneling, dissipation, and interactions.

\subsection{Spectral sensitivity based on the scalings of complex pseudospectra}

In this paragraph, we are going to characterize the sensitivity of the open Bose-Hubbard dimer, by relying on the associated Liouvillian pseudospectra scalings \cite{Komis2022}. More specifically, to characterize the associated spectral sensitivity, we compute the complex pseudospectrum scaling with the perturbation strength for given dissipation strengths. In particular, in Fig.~\ref{fig:spectral_analysis_BH}(c), we see how the scalings of the complex pseudospectra reveal the order of the underlying higher-order EPs. More specifically, two
EPs of second order and one EP of third order are evident. Similarly, in Figs.~\ref{fig:SpectralU}(b) and \ref{fig:SpectralU}(c), we show the pseudospectrum at $\gamma \approx 1.625$ and $\gamma = 2$, respectively. At $\gamma \approx 1.625$, the pseudospectrum reveals eight EP2 degeneracies and a single EP3 degeneracy, manifested through elongated and overlapping pseudospectral eigenvalue clouds. At $\gamma = 2$, eight EP2 degeneracies (two of them at $\lambda=-1/2$) and one EP3 are present.

\begin{figure}
\includegraphics[width=1.0\linewidth]{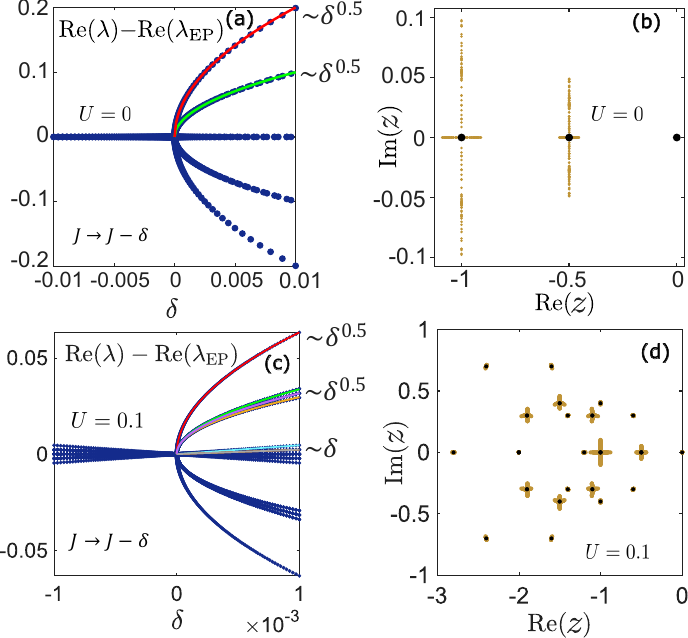}
\caption{Hopping-induced sensitivity and structured pseudospectra near exceptional points for $\gamma = 2$.
(a) Real part of Liouvillian eigenvalues minus their EP value, as a function of the perturbation strength $\delta$ for the noninteracting case. The perturbation is implemented by varying the hopping amplitude as $J \rightarrow J - \delta$. Blue dots denote numerical data, while solid lines represent power-law fits to these data. (b) Corresponding structured (parameter-dependent) pseudospectrum at the EP for the noninteracting case, illustrating spectral sensitivity under restricted perturbations of the hopping parameter ($J \rightarrow J - \delta \xi$, where $\xi$ is a normal random number and $\delta=10^{-3}$). Same analysis for the interacting case: eigenvalue splitting [panel (c)] and the associated structured pseudospectrum at the EP [panel (d)]. 
}
\label{fig:sensitivity}
\end{figure}

\subsection{Spectral sensitivity based on structured pseudospectra}

While the complex pseudospectrum provides a global measure of non-Hermitian sensitivity under arbitrary perturbations, physically relevant perturbations are those related to actual system parameters. We therefore complement the complex pseudospectra with structured pseudospectra, where perturbations are applied only to particular physical parameters of the Bose-Hubbard model.

We first consider perturbations of the hopping amplitude, implemented as $J \rightarrow J - \delta$, where $\delta$ controls the perturbation strength. Figure~\ref{fig:sensitivity}(a) shows the resulting splitting of the real parts of the Liouvillian eigenvalues relative to their EP value for the noninteracting case. The eigenvalue shifts exhibit clear power-law scaling with $\delta$, in agreement with the EP2s extracted from the complex pseudospectrum. The corresponding structured  pseudospectrum, shown in Fig.~\ref{fig:sensitivity}(b), visualizes the enhanced sensitivity under restricted perturbations of the hopping parameter. Figures~\ref{fig:sensitivity}(c) and \ref{fig:sensitivity}(d) present the same analysis for the interacting case. Interactions lead to a richer sensitivity landscape, with multiple EPs contributing to the observed scaling behavior. Nevertheless, the parameter-induced eigenvalue splitting continues to follow power laws consistent with the underlying EP structure, demonstrating that structured sensitivity faithfully captures the non-Hermitian response even in the presence of interactions.

\begin{figure}
\includegraphics[width=1.0\linewidth]{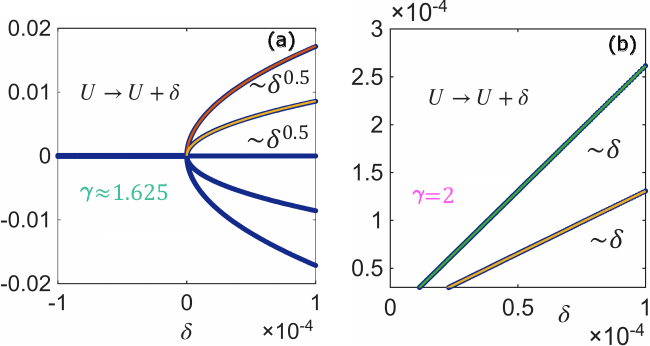}
\caption{Interaction strength-induced sensitivity near exceptional points.
Real part of Liouvillian eigenvalues minus their EP value as a function of the perturbation strength $\delta$, where the interaction strength is varied as $U \rightarrow U + \delta$. Blue dots denote numerical data, while solid lines represent power-law fits to these data. (a) For EP at $\gamma \approx 1.625$. (b) For EP at $\gamma = 2$. { In panel (b), only representative segments are shown for illustrative purposes.}}
\label{fig:sensitivityU}
\end{figure}

Finally, we examine the sensitivity of the spectrum to perturbations in the interaction strength, $U \rightarrow U + \delta$, as shown in Fig.~\ref{fig:sensitivityU}. The resulting eigenvalue shifts again exhibit clear power-law scaling near the EPs, with exponents that depend on the dissipation strength $\gamma$. At $\gamma \approx 1.625$, the presence of EP2s is reflected in the corresponding scaling behavior. At $\gamma = 4J$, the EPs are inherited from the one-boson sector, where the interaction term vanishes. The branches shown in Fig.~\ref{fig:sensitivityU}(b) shift linearly with $\delta$ to leading order.

Taken together, these results establish structured sensitivity analysis as a powerful and experimentally relevant probe of EP physics. By restricting perturbations to tunneling and interaction strengths, one can directly observe EP-induced scaling behavior without relying on fully unstructured perturbations. This perspective bridges the gap between complex pseudospectrum and structured pseudospectra in realistic open quantum systems.

\subsection{Dynamics and long-time averages}

\begin{figure}
\includegraphics[width=1\linewidth]{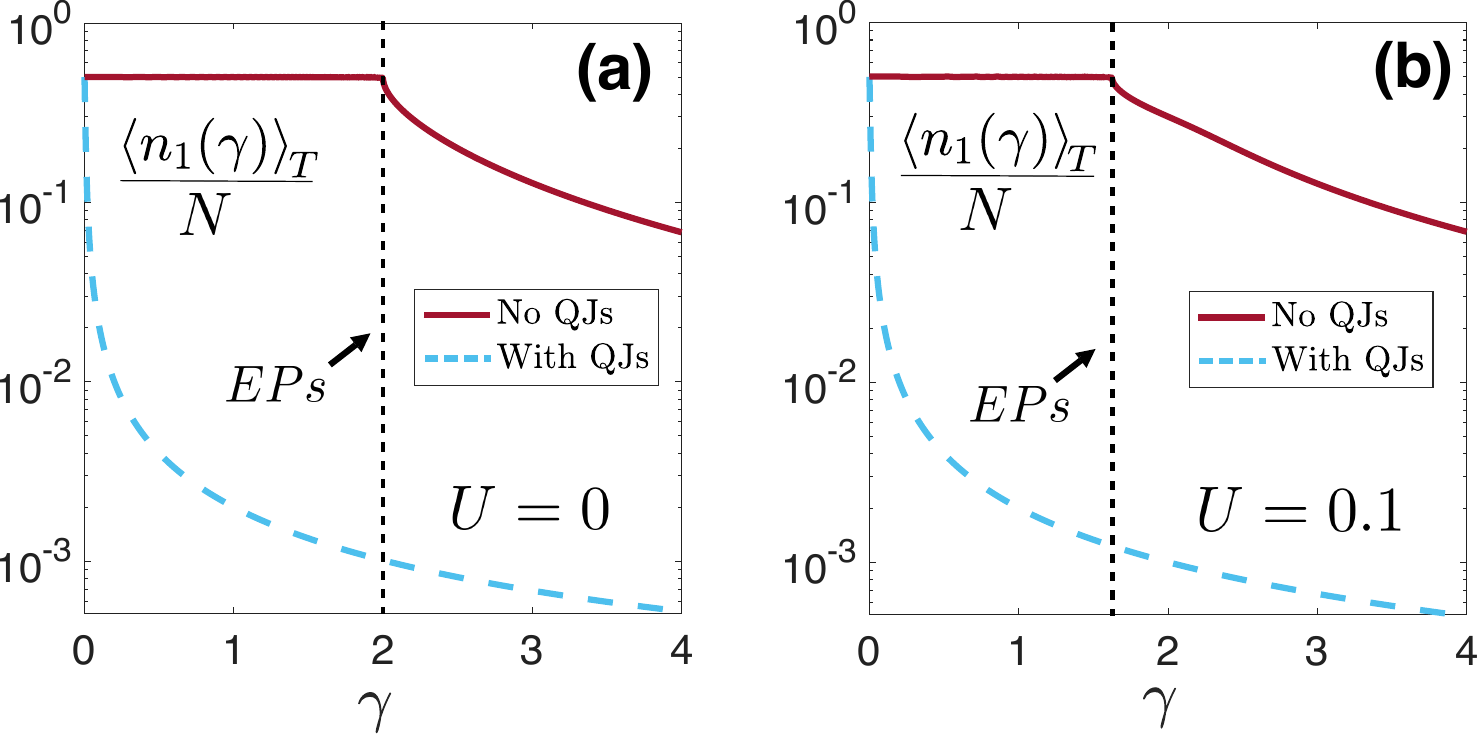}
\caption{Long-time averaged population dynamics in the Bose-Hubbard model.
LTA of the site occupation $\langle n_1(\gamma) \rangle_T$ as a function of the dissipation rate $\gamma$. Dashed blue lines correspond to the full Lindblad dynamics with QJs,
while solid red lines show the evolution obtained by omitting QJs. Including QJs leads to smooth decay as $\gamma$ increases, whereas omitting QJs we see an oscillatory phase before the EP, followed by decay after it. (a) Noninteracting case. (b) Interacting case. The population average is normalized by the initial particle number. 
}
\label{fig:dynamics_longtime_BH}
\end{figure}

Now, we continue with the investigation of possible signatures of EPs on the quantum dynamics, when the system operates around these singularities. In particular, we examine two different cases: when the QJs are omitted from Eq.~\eqref{master_equation}, and when they are included. For the first case, to calculate expectation values, as mentioned, we have to divide by the trace of the reduced density matrix or, in other words, use Eq.~\eqref{expectationnoQJ}. For more details, see Appendix \ref{app:linbland}. Moreover, the two observables that we focus on are the populations on the first site and the population imbalances. 

Regarding the first one, the LTA of the population of the first site of the Bose-Hubbard system (denoted as $\langle n_1(\gamma) \rangle_T$ and normalized with the initial particle number $N$) is shown in Fig. \ref{fig:dynamics_longtime_BH}(a), highlighting the fundamental distinctions between the cases with and without QJs. In the no-QJ case (solid red line), the LTA highlights a sharp transition at the EP. Before the EP, the oscillatory dynamics lead to a relatively high LTA (see also Fig. \ref{fig:Dynamics_BH} in Appendix \ref{app:Dyn}). At the EP, the transition is abrupt, with oscillations rapidly suppressed and the system entering a steady-state regime. This sharp change serves as a clear marker of the EP, making the LTA a useful diagnostic tool for identifying critical points in the parameter space. This case is practically identical to the case where the study is performed with the effective NHH (see Fig. \ref{fig:h_effBH} in Appendix \ref{app:eff}).

For the case with QJs (blue dashed line), the LTA is smoother, consistent with dissipative dynamics. Unlike the sharp transitions observed in the no-QJ case, the inclusion of QJs ensures that the system evolves smoothly across all dissipation rates. The difference in the dynamics is illustrated with the help of the QJM (see Fig. \ref{fig:qj_BH} in Appendix \ref{app:qjm}).

One thing to note here is that, if we calculate $\langle n_1(\gamma) \rangle_T$ analytically [see Eq. \eqref{exact_lta_noQJs} and Appendix \ref{app:bose} for details on the derivation], we see that if we consider perturbations of the hopping amplitude, implemented as $J \rightarrow J - \delta$ as in the previous section, the response is [see also Eq. \eqref{exact_lta_noQJs2}]
\begin{equation}
\frac{1}{2} - \langle n_1(\gamma) \rangle_T \propto \sqrt{\delta},
\end{equation}
signaling a case where we can infer the order of the EP from the dynamics.

The interacting case, shown in Fig.~\ref{fig:dynamics_longtime_BH}(b), exhibits similar qualitative features. Without QJs (solid red line), the system undergoes a sharp transition at the EPs at $\gamma \approx 1.625$. A few remarks are warranted regarding the case without QJs. The nature of the dynamics depends on the choice of initial conditions: If the initial state includes two-particle Fock states or their superpositions, the transition occurs near $\gamma \approx 1.625$; only if single-particle initial states are used, the transition takes place at $\gamma = 2$. The physical picture remains the same; thus, these results confirm that interactions do not alter the fundamental impact of QJs on the system's dynamics. The aforementioned physical picture also holds for larger systems (see Fig. \ref{fig:3bosons} in Appendix \ref{app:large}).

\begin{figure}
\includegraphics[width=1.0\linewidth]{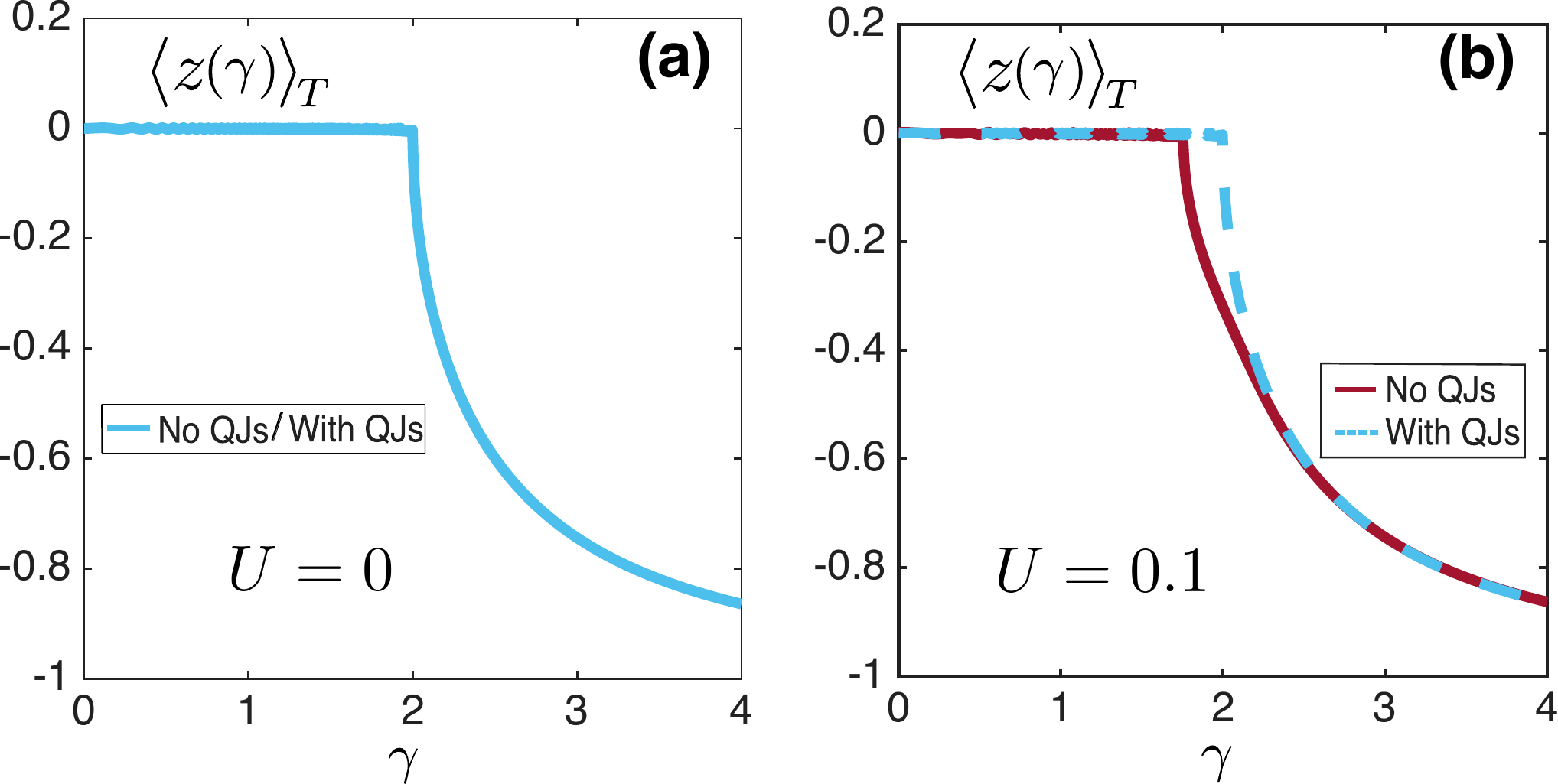}
\caption{Long-time averaged imbalance dynamics in the Bose-Hubbard model as a function of the dissipation rate $\gamma$. (a) Noninteracting case showing the same behavior and the same cusp regardless of whether we include the QJs or not (analytical result). (b) Interacting case. We observe different transition points between oscillatory and dissipative regimes depending on whether QJs are included.}
\label{fig:imbalance}
\end{figure}

Regarding the second observable, we consider the population imbalance. In this case, such an experimentally relevant quantity exhibits change around the EP, even in the full Liouvillian case. The population imbalance is defined in the simplest case for the Bose-Hubbard dimer as
\begin{equation}
\label{imbalance_definition}
\langle {z (t)} \rangle = \frac{\langle {n_1 (t)} \rangle-\langle {n_2 (t)} \rangle}{\langle {n_1 (t)} \rangle+\langle {n_2 (t)} \rangle}.
\end{equation}
This quantity is typically studied in bosonic Josephson junctions \cite{Albiez2005} and other Bose-Hubbard dimer systems. The LTA of the imbalance, denoted as $\langle z(\gamma) \rangle_T$, is plotted in Fig. \ref{fig:imbalance} for both the interacting and the noninteracting case. For the noninteracting case, the result with and without QJs is the same [for the analytical result, see Eq. \eqref{exact_imbalance}]. Moreover, in the interacting case, the dynamics still changes in a similar manner as in the population dynamics but at a different EP, depending on whether we include the QJs or not.

One interesting aspect is that once again we can find the order of the EP from the imbalance dynamics. The LTA $\langle z(\gamma) \rangle_T$ can be calculated analytically in the noninteracting case [see Eq. \eqref{lta_imbalance}] and the response to the same perturbation as before is [see also Eq. \eqref{lta_imbalance_response}]
\begin{equation}
  \left|\langle z(\gamma) \rangle_T\right| \propto \sqrt{\delta}.  
\end{equation}

In summary, the Bose-Hubbard dimer provides a minimal yet rich setting for exploring the interplay between dissipation, QJs, and EPs. While the Liouvillian eigenvalue spectrum is insensitive to the inclusion of QJs, both spectral sensitivity and dynamical observables reveal pronounced EP signatures. Complex and structured pseudospectral analyses expose multiple EP2 and EP3 degeneracies and quantify their associated sensitivity, while parameter-dependent perturbations of tunneling and interactions demonstrate experimentally relevant EP-induced scaling. At the dynamical level, LTA of the population and imbalance observables distinguish sharply between conditional (no-QJ) and unconditional (full Lindblad) evolution, showing that EP-induced transitions are smoothed by QJs yet remain accessible through suitably normalized quantities. Together, these results establish the Bose-Hubbard model as a benchmark for understanding how EP sensitivity and dynamics manifest—and differ—in open quantum systems.

\section{Non-Hermitian fermionic system: Open spin chain}
\label{spin_chain}

\begin{figure*}
\centering
\includegraphics[width=0.75\linewidth]{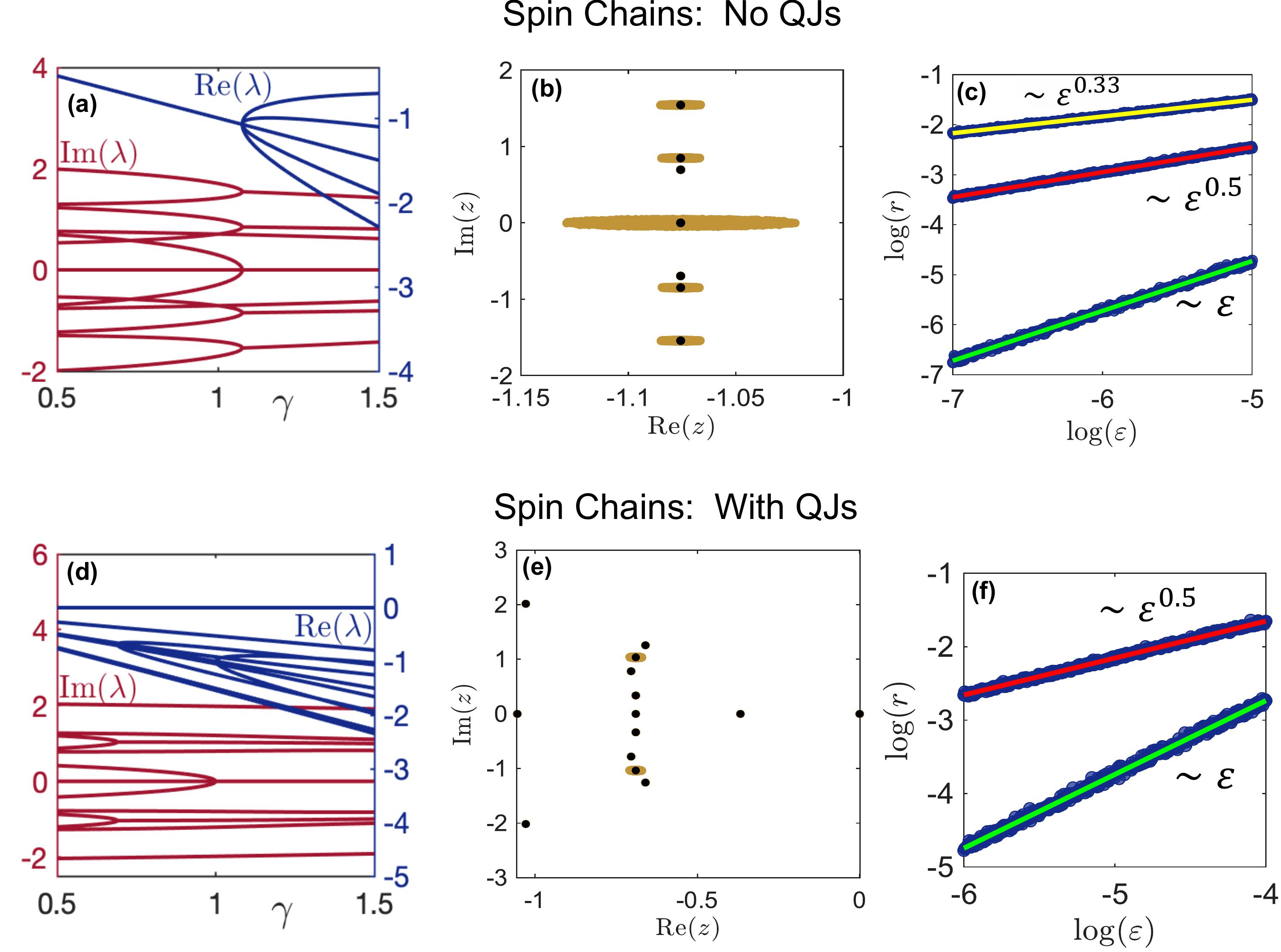}
\caption{Liouvillian eigenspectrum and associated complex pseudospectra of the spin chain model.
Top row: results obtained without QJs.
Bottom row: results including QJs.
(a), (d) Real and imaginary parts of the Liouvillian eigenvalues as a function of the dissipation rate $\gamma$,
revealing EPs at $\gamma \approx 1.076$ without QJs and at $\gamma \approx 0.689$ and $\gamma = 1$ with QJs.
(b), (e) Complex pseudospectra at the corresponding EPs, illustrating enhanced spectral sensitivity.
Black dots denote eigenvalues, while colored dots show the  pseudospectral eigenvalue clouds ($\epsilon = 10^{-5}$ and $10^{-3}$, respectively).
(c), (f) Scaling of the pseudospectral radius $r$ with perturbation strength $\epsilon$. Blue dots denote numerical data. The fitted scaling exponents in panel (c) are approximately $1$ (green), $0.5$ (red), and $0.33$ (yellow). In panel (f), they are approximately $1$ (green) and $0.5$ (red). Without QJs [panel (c)], the scaling reveals four EP2s and one EP3, whereas with QJs [panel (f)] only two EP2s are present.
For $\gamma = 1$, the pseudospectrum is omitted for clarity, although an EP2 is found.
}
\label{fig:spectral_analysis}
\end{figure*}

Spin chains have long been used to explore open-system effects in quantum magnetism, many-body quantum dynamics, and nonequilibrium transport, particularly in the presence of dissipation \cite{Breuer2007,Bose2003}. They provide an ideal framework for studying how coherent and dissipative interactions compete to shape the system's evolution.
Beyond their role in condensed matter physics, they are also key models in quantum information science since they serve as natural platforms for quantum state transfer, where spins can encode and transmit quantum information with high fidelity \cite{Bose2003,Christandl2004}. Additionally, they are fundamental in quantum computation, as spin-based systems can implement quantum gates and error-correcting codes \cite{Briegel2009,Ladd2010}. In the context of open quantum systems, dissipative spin chains have been proposed as reservoir engineering schemes for robust entanglement generation \cite{Verstraete2009}. 

Here, we focus primarily on a  two-spin system, since it is a minimal model where non-Hermitian spectral properties and QJ effects can be examined in a systematic manner. In particular, the Hamiltonian of the two-spin chain system is \cite{Minganti2019,Kulkarni2023}
\begin{align}
    {H}_{\text{SC}} &= \frac{\Omega}{2} \left( \sigma_x \otimes I_2 + I_2 \otimes \sigma_x \right) + \Delta \left(\sigma_z \otimes I_2 + I_2 \otimes \sigma_z \right)\nonumber \\
      &\quad - J \sigma_z \otimes \sigma_z.
\end{align}
where \( \sigma_x \) and \( \sigma_z \) are the Pauli matrices, while \( I_2 \) is the \( 2 \times 2 \) identity matrix. $\Omega$ is the Rabi frequency, $\Delta$ is the detuning, and $J$ is the interaction strength. An operator acting on a specific spin, such as \( \sigma_{z_1} \), is formally defined as \( \sigma_{z_1} = \sigma_z \otimes I_2 \), while \( \sigma_{z_2} = I_2 \otimes \sigma_z \). This ensures that the operator acts only on the corresponding spin subspace.

Interaction with the environment is described by the Lindblad operators
\begin{equation}
    L_1 = \sqrt{\gamma_1}\,\sigma_1^-, \qquad
    L_2 = \sqrt{\gamma_2}\,\sigma_2^+ ,
\end{equation}
representing loss on the first spin and gain on the second spin.
We consider the balanced gain-loss case $\gamma_1=\gamma_2=\gamma$ and comment briefly on the unbalanced case when considered relevant. In all simulations, the system is initialized with both spins in the $\ket{\uparrow}$ state. In a closed spin chain system with \( N \) spins, the Hilbert space dimension is \( 2^N \), as each spin has two possible states (e.g., ``up" and ``down"). In an open system described by a density matrix, the Liouvillian acts on the space of operators, leading to a dimension of \( 2^{2N} \) for the Liouvillian superoperator. Unless otherwise specified, we fix $\Omega=1$, $\Delta=0$, and $J=1/4$. 
In the following two subsections, we summarize the spectral, sensitivity, and dynamical properties of this model, emphasizing how they corroborate and extend the results obtained for the Bose-Hubbard system.

\subsection{Liouvillian eigenspectra, complex
pseudospectra, structured perturbations}

We now briefly discuss the spectral and pseudospectral properties of the two-spin Liouvillian, focusing on the qualitative role played by QJs. Figure~\ref{fig:spectral_analysis} shows the Liouvillian spectrum, complex pseudospectrum, and corresponding scalings, both without (top row) and with (bottom row) QJs. When QJs are omitted [Fig.~\ref{fig:spectral_analysis}(a)], the Liouvillian spectrum exhibits multiple eigenvalue coalescences at
{ $\gamma \approx 1.076$}, signaling the presence of several EPs. These eigenvalue bifurcations coincide with those obtained from the effective NHH, as discussed in Appendix~\ref{app:eff}. In contrast, when QJs are included [Fig.~\ref{fig:spectral_analysis}(d)], the Liouvillian spectrum itself is modified: The EPs are shifted to $\gamma \approx 0.689$ and $\gamma = 1$, demonstrating that the full Lindblad dynamics qualitatively alter the non-Hermitian spectral structure. This behavior stands in sharp contrast to the Bose-Hubbard model, where the Liouvillian spectrum is insensitive to the inclusion of QJs because the Hamiltonian conserves the particle number and the losses only lower it.

Further insight is obtained from the complex pseudospectrum, shown in Figs.~\ref{fig:spectral_analysis}(b) and \ref{fig:spectral_analysis}(e). In both cases, the pseudospectrum reveals strong non-Hermitian sensitivity near the EPs,
but its structure differs markedly depending on whether QJs are included. This is quantified by the scaling of the pseudospectral radius $r$ with perturbation strength $\epsilon$, shown in Figs.~\ref{fig:spectral_analysis}(c) and \ref{fig:spectral_analysis}(f). Without QJs, the scaling reveals four EP2 degeneracies and one EP3, whereas including QJs reduces the EP structure to two EP2s. Thus, QJs do not merely shift EP locations but also lower the effective EP order and suppress higher-order degeneracies.

\begin{figure}
\includegraphics[width=1.0\linewidth]{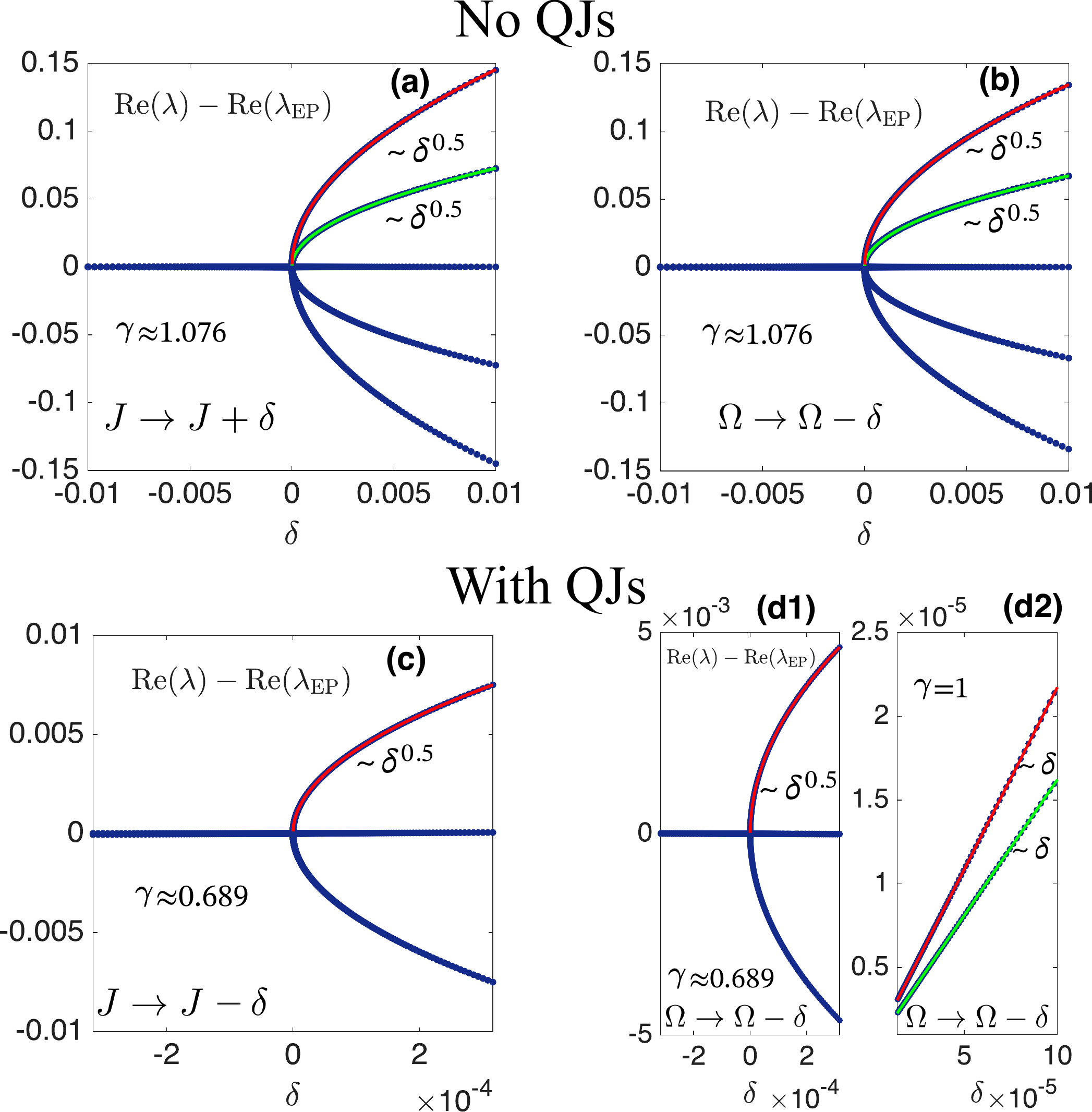}
\caption{Spectral sensitivity in the two-spin system. { Real part of Liouvillian eigenvalues minus their EP value as a function of the perturbation strength $\delta$. The direction of each structured perturbation is indicated directly inside the corresponding panel. Panels (a) and (b) show the sensitivity without QJs at the EP $\gamma \approx 1.076$. Panels (c) and (d1) show the corresponding results with QJs included at $\gamma \approx 0.689$, while panel (d2) shows the Rabi-frequency perturbation at $\gamma=1$. Blue dots denote numerical data, while solid lines represent power-law fits to these data. Square-root scaling $\sim \delta^{1/2}$ signals EP2 behavior, except in panel (d2), where linear scaling is observed.}
}
\label{fig:sensitivitySC}
\end{figure}

Moreover, we characterize the non-Hermitian response of the two-spin system through a structured sensitivity analysis,
shown in Fig.~\ref{fig:sensitivitySC}. Parameter-induced spectral sensitivity is probed by perturbing either the interaction strength or the Rabi frequency and monitoring the resulting splitting of the real parts of the Liouvillian eigenvalues relative to their EP value. Both without and with QJs, most cases exhibit square-root scaling, $\mathrm{Re}(\lambda)-\mathrm{Re}(\lambda_{\mathrm{EP}}) \sim \delta^{1/2}$,
consistent with EP2 behavior. { For the Rabi-frequency perturbation at $\gamma=1$, shown in Fig.~\ref{fig:sensitivitySC}(d2), the response is instead linear in $\delta$.} Overall, these results show that while QJs modify the Liouvillian spectrum, EP-induced parameter sensitivity persists in the full Lindblad description whenever an EP is present.

\subsection{Dynamics and long-time averages}

\begin{figure}
\includegraphics[width=0.5\linewidth]{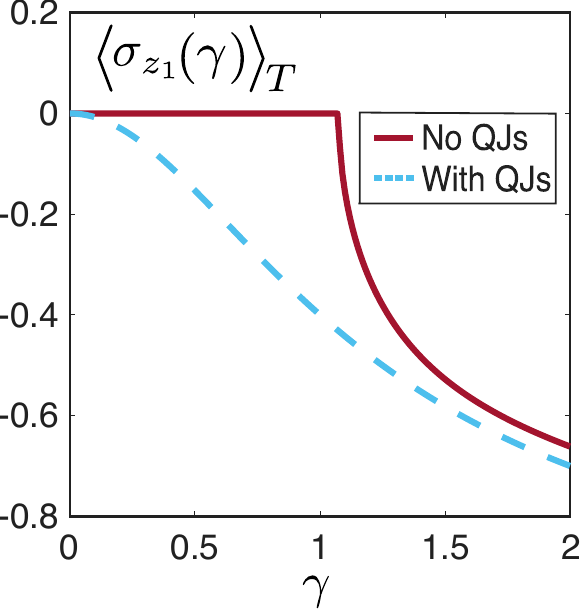}
\caption{LTA of the observable \( \langle \sigma_{z_1} \rangle \) as a function of \( \gamma \). With QJs (dashed blue line) we only have dissipative dynamics, while without QJs (solid red line) we have oscillatory dynamics until the EP at { $\gamma \approx 1.076$} and dissipation to a steady state afterward (see also Fig. \ref{fig:Dynamics_SC}).}
\label{fig:dynamics_longtime}
\end{figure}

In this paragraph, we examine the LTA $\langle \sigma_{z_1}(\gamma) \rangle_T$, and our results are shown in Fig.~\ref{fig:dynamics_longtime}. It has a similar behavior to the population in the Bose-Hubbard case. When QJs are omitted (solid red line), the dynamics are governed by the effective NHH and remain oscillatory for dissipation strengths below the EP, resulting in a constant LTA. At the EP, these oscillations are abruptly suppressed, leading to a sharp transition to a dissipative steady state. In contrast, when QJs are included (dashed blue line), the system dissipates across the entire parameter range. No sharp dynamical signature is observed at the EP; instead, the LTA evolves continuously toward its steady-state value. Beyond the EP, both descriptions approach a steady state. Finally, we note that omitting QJs yields results that are practically identical to those obtained from the effective non-Hermitian Hamiltonian (see Appendix~\ref{app:eff}).

To further investigate the role of QJs in the dynamics of the two-spin system, we employ the QJM  to simulate individual trajectories and compare them to the exact Liouvillian evolution. Our results are summarized in Appendix \ref{app:qjm} and are similar to the ones for the Bose-Hubbard case. { We have also checked the robustness of the results for larger systems. In particular, we have calculated the LTA for spin chains with three and four spins, shown in Fig.~\ref{fig:spins34} of Appendix~\ref{app:large}. The results display the same qualitative distinction between the no-QJ and full Lindblad dynamics.}

\section{Conclusions and perspectives}
\label{conper}

In this work, we have investigated the interplay between EPs and QJs, and their effects on spectral sensitivity and dynamics in open quantum systems, focusing on two paradigmatic models: a two-site Bose-Hubbard dimer and a dissipative two-spin system. { The two-spin model is used as the minimal spin-chain example, supplemented by finite-size results for three and four spins in Appendix~\ref{app:large}.} Using the Lindblad master equation, we systematically compared descriptions that omit QJs, equivalent to effective non-Hermitian Hamiltonian dynamics, with the full Lindbladian evolution that includes stochastic jump processes.

Our analysis demonstrates that QJs play a crucial role in reshaping both the spectral and dynamical manifestations of EP physics. In the dissipative Bose-Hubbard dimer, the spectrum is the same whether we include the QJs or not; however, { in the driven-dissipative spin systems considered here, where the drive does not conserve the number of excitations and gain is also present,} the spectrum is different. Using the complex pseudospectrum, we found the orders of the EPs. Structured sensitivity analysis showed that EP-induced power-law scaling of eigenvalue splittings remains observable whenever an EP is present in most of the cases considered.

On the dynamical level, we showed that QJs qualitatively modify the signatures of EPs in time evolution. When QJs are omitted, coherent oscillations dominate below the EP and are abruptly suppressed at the EP, producing sharp dynamical transitions that can be directly inferred from LTA of observables. In contrast, the inclusion of QJs leads to dissipative dynamics toward steady states across all dissipation strengths, possibly through damped oscillations, thereby smoothing out critical dynamical behavior. Despite this smoothing, suitably chosen observables—particularly ratio-based quantities such as population imbalance—retain clear sensitivity to EPs, even in the full Lindblad description.

Overall, our results lead to two main conclusions: First, the Liouvillian pseudospectra (complex and structured) characterize the spectral sensitivity of our open quantum systems, and second, signatures of the Liouvillian EPs on dynamics are evident when we consider long-time averages of populations or population imbalances. Our investigation highlights the limitations of effective non-Hermitian approaches for predicting stability and critical behavior, and emphasizes the necessity of full Lindbladian modeling when QJs are experimentally relevant.

Several directions for future work naturally follow. Extending the present analysis to larger many-body systems would enable the exploration of the interplay between EPs, interactions, and collective dissipation beyond minimal models. A particularly promising avenue lies in the context of quantum informatics: Investigating entanglement measures and higher-order correlations could elucidate how EPs and QJs influence quantum correlations and information-theoretic resources in open systems. On the spectral side, structured pseudospectra analysis, especially under experimentally constrained perturbations, may provide a relevant method for diagnosing stability and sensitivity in noisy non-Hermitian platforms. Finally, experimental implementations in systems such as cold atoms, photonic lattices, and superconducting circuits offer promising routes to directly test the predictions presented here and to exploit EP physics in realistic open quantum devices.

\begin{acknowledgments}
The authors acknowledge financial support from the European Research Council (ERC) through the Consolidator Grant Agreement No. 101045135 (Beyond\_Anderson). G.K.S. further acknowledges funding from the Stavros Niarchos Foundation (SNF) and the Hellenic Foundation for Research and Innovation (H.F.R.I.) under the 5th Call of the “Science and Society” Action, titled “Always strive for excellence---Theodoros Papazoglou” (Project No. 11496, “PSEUDOTOPPOS”). M.K. is grateful for the hospitality of the Department
of Physics, University of Crete (UOC), and the Institute of Electronic Structure and Laser (IESL)—FORTH,
at Heraklion, Greece. M.K. acknowledges support from the Department of Atomic Energy, Government of India, under Project No. RTI4001.
\end{acknowledgments}

\section*{Data availability}
The data that support the findings of this article are not publicly available. The data are available from the authors upon reasonable request.

\appendix

\section{Solution of the Lindblad equation}
\label{app:linbland}

For numerical calculations, it is convenient to rewrite the Lindblad master equation \eqref{master_equation} in matrix form by vectorizing the density matrix $\rho(t)$. Vectorization maps a $d \times d$ matrix, where $d$ is the Hilbert-space dimension, to a $d^2$-dimensional column vector $\mathrm{vec}(\rho(t))$ by stacking its columns sequentially.

Under this transformation, the Lindblad equation becomes a first-order linear differential equation,
\begin{equation}
\frac{d}{dt}\mathrm{vec}(\rho(t)) = \mathcal{L}\,\mathrm{vec}(\rho(t)),
\end{equation}
where $\mathcal{L}$ is the Liouvillian superoperator—a $d^2 \times d^2$ matrix encoding both coherent and dissipative contributions:
\begin{align}
\mathcal{L} &= -\frac{i}{\hbar} (I \otimes H - H^T \otimes I) \nonumber \\
&\quad + \sum_k \left(
L_k^* \otimes L_k
- \frac{1}{2}
\big(
I \otimes L_k^\dagger L_k
+ (L_k^\dagger L_k)^T \otimes I
\big)
\right).
\end{align}
The first term describes unitary evolution generated by the Hamiltonian, while the remaining terms account for dissipative processes induced by the Lindblad operators.

For a time-independent Liouvillian, the formal solution is
\begin{equation}
\mathrm{vec}(\rho(t)) = e^{\mathcal{L}t}\,\mathrm{vec}(\rho(0)),
\end{equation}
where $e^{\mathcal{L}t}$ denotes the matrix exponential. When an analytical expression is unavailable, the evolution can be obtained either by numerical evaluation of the matrix exponential or by integrating the differential equation using standard ordinary differential equation solvers. In this work, we employ direct numerical evaluation of the matrix exponential.

Once $\mathrm{vec}(\rho(t))$ is obtained, the density matrix $\rho(t)$ is reconstructed by reshaping the vector back into a $d \times d$ matrix. Expectation values of observables are then computed using Eq.~\eqref{expectation} for the full Lindblad dynamics. When QJs are omitted, the trace of $\rho(t)$ decays under the non-Hermitian evolution generated by $H_{\text{eff}}$, and normalized expectation values \eqref{expectationnoQJ} must be used.

This vectorization framework provides a robust and general method for computing the dynamics, steady states, and spectra of open quantum systems. It also forms the basis for analytical treatments when the Liouvillian can be diagonalized explicitly, as discussed in Appendix~\ref{app:bose}.

\section{Results using the effective Hamiltonian}
\label{app:eff}

\begin{figure*}
\includegraphics[width=0.75\linewidth]{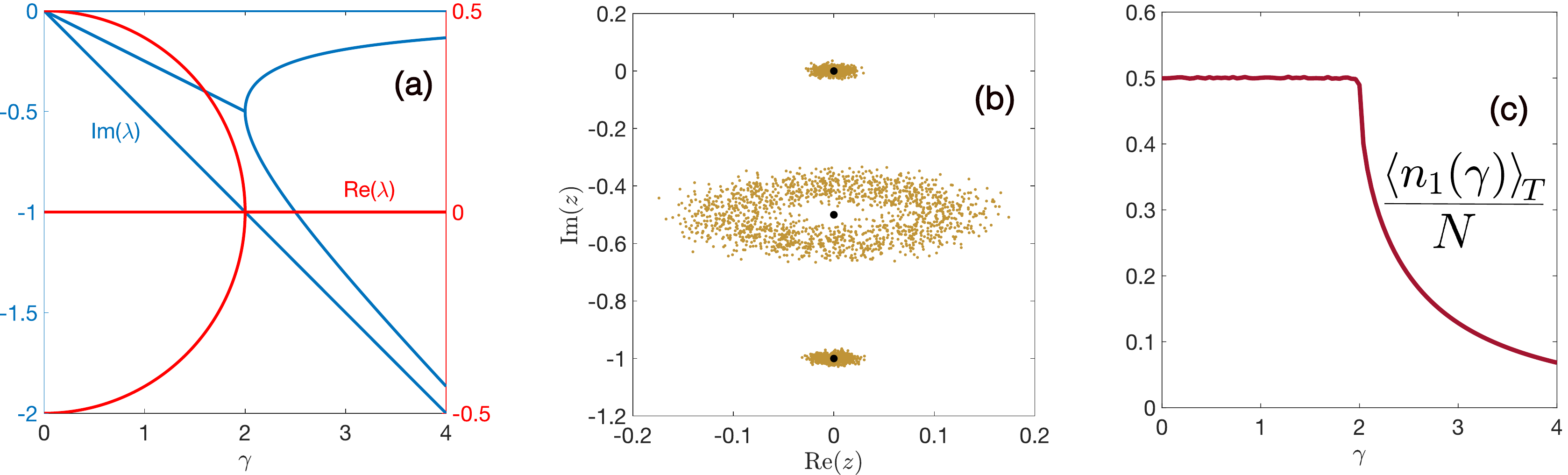}
\caption{Results of the Bose-Hubbard system using the effective NHH in the noninteracting case. (a) Spectrum showing EP at $\gamma = 4J$ (analytical result). (b) Pseudospectrum at this value of $\gamma$  ($\epsilon = 10^{-5}$).  (c) LTA $ \langle n_1 (\gamma) \rangle_T $ as a function of $\gamma$. Parameters are the same as in the main text.}
\label{fig:h_effBH}
\end{figure*}

\begin{figure*}
\includegraphics[width=0.75\linewidth]{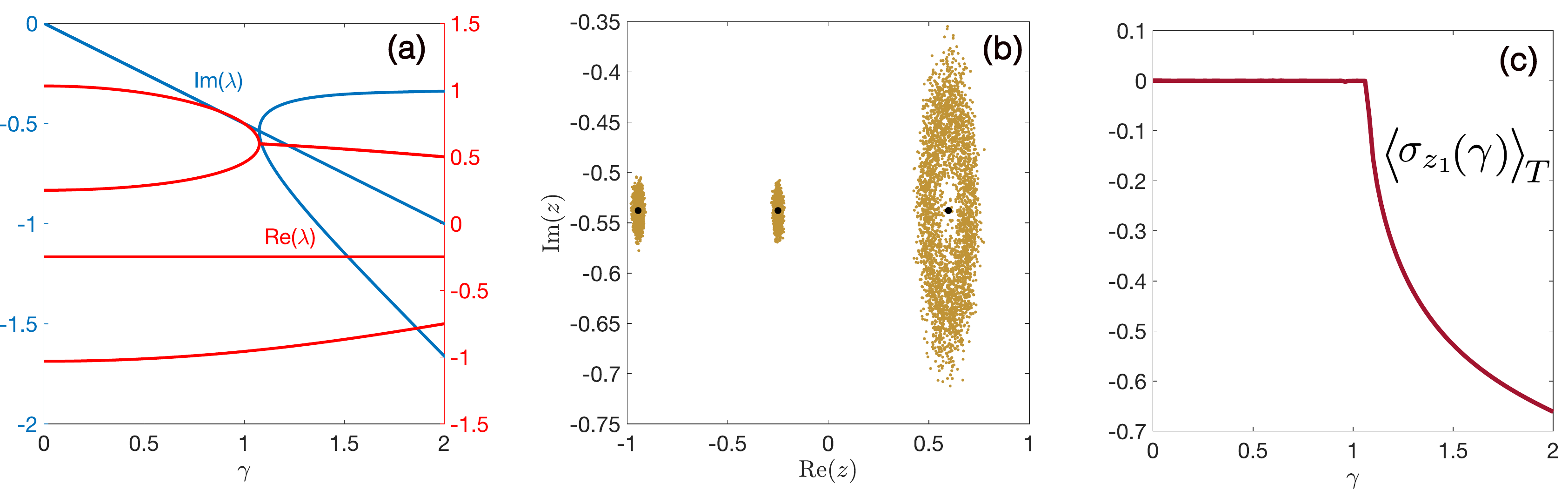}
\caption{Results of the two-spin chain system using the effective NHH. (a) Spectrum showing EP at { $\gamma \approx 1.076$}, same as the ``no QJs" case presented in the main text. (b) Pseudospectrum at this value of $\gamma$ ($\epsilon = 10^{-5}$). (c) LTA \( \langle \sigma_{z_1} (\gamma) \rangle_T \) as a function of $\gamma$. Parameters are the same as in the main text. 
}
\label{fig:h_eff}
\end{figure*}

In this Appendix, we present spectral and dynamical results obtained using the effective non-Hermitian Hamiltonian $H_{\mathrm{eff}}$ defined in Eq.~\eqref{nhh}. For both the Bose-Hubbard dimer (Fig. \ref{fig:h_effBH} ) and the two-spin system (Fig. \ref{fig:h_eff}), the resulting spectra, pseudospectra, and LTA are similar to those obtained from the Lindblad equation when QJs are omitted. This comparison confirms that neglecting QJs leads to an effective description governed by $H_{\mathrm{eff}}$, while highlighting the limitations of such an approach when full Lindblad dynamics are considered.

\section{Pseudospectra and related metrics}
\label{app:pseudospectrum}

In this Appendix, we discuss the concept of pseudospectrum \cite{Trefethen2005} of a non-Hermitian matrix and explain its numerical implementation, as used in our analysis of EPs. The pseudospectrum quantifies spectral sensitivity and allows one to extract the order of EPs through a characteristic power-law scaling with perturbation strength. In this work, we use the perturbation-based definition of the pseudospectrum. The \emph{complex pseudospectrum} \( \Lambda_\epsilon(A) \) of a square matrix \( A \in \mathbb{C}^{d \times d} \) is defined as the set of complex numbers \( z \in \mathbb{C} \) satisfying the condition
\begin{equation}
\begin{split}
\Lambda_\epsilon(A)
= \Big\{\, z \in \mathbb{C} \;\Big|\;
&z \text{ is an eigenvalue of } A + E,\\
&\text{with } \|E\| < \epsilon \,\Big\},
\end{split}
\end{equation}
which interprets the pseudospectrum as the set of all possible eigenvalues resulting from perturbations of \( A \) with bounded norm.

Thus, we generate perturbations of the form
\begin{equation}
\label{ps3}
E = N \frac{\mathcal{E}}{||\mathcal{E}||},
\end{equation}
where \( \mathcal{E} \) is a random complex matrix whose real and imaginary parts follow a normal distribution, $||...||$ is the matrix two-norm, and \( N = \sqrt{d/2} \) is a normalization factor ensuring consistent scaling across different system sizes \cite{Komis2022}.

To quantify the extent of the pseudospectrum, we define its \emph{radius} \( r_\epsilon(A) \) as the maximal distance of any point in \( \Lambda_\epsilon(A) \) from the origin:
\begin{equation}
\label{rad}
r_\epsilon(A) = \sup_{z \in \Lambda_\epsilon(A)} |z|.
\end{equation}

As the system approaches an EP, the pseudospectrum exhibits a characteristic scaling behavior. In particular, the radius \( r \) scales with the perturbation strength \( \epsilon \) as
\begin{equation}
r \propto \epsilon^{1/n},
\end{equation}
where \( n \) is the order of the EP \cite{Wiersig2014}. This power-law scaling enables the determination of EP order by fitting the dependence of \( r \) on \( \epsilon \).

In many physical settings, perturbations are not arbitrary but have a specific structure, reflecting how parameters enter the Hamiltonian. For example, in the Bose-Hubbard model, the coupling strength \( J \) may be experimentally or numerically tuned, while other matrix elements remain fixed. To capture such cases, we employ the \emph{structured pseudospectrum} \cite{Kiorpelidis2026}. In our analysis, we consider perturbations of the hopping amplitude
\begin{equation}
J \rightarrow J - \epsilon \xi,
\end{equation}
where $\xi$ is a normally distributed random number, and compute the corresponding eigenvalues of \( A(J - \epsilon\xi) \).  The resulting structured pseudospectrum highlights how eigenvalues move in the complex plane under controlled parameter changes, in contrast to random unstructured perturbations. The structured pseudospectrum provides a clearer connection between spectral sensitivity and physically accessible tuning parameters.

\section{Analytical results in the Bose-Hubbard model}
\label{app:bose}

For the Bose-Hubbard model defined in Eq.~\eqref{BHH}, the noninteracting case ($U=0$) is exactly solvable for arbitrary particle number $N$ and lattice size $M$. The system reduces to a quadratic bosonic problem that can be diagonalized analytically. Here, we present explicit results for the two-site, single-particle case discussed in Sec.~\ref{bose_hubbard}, with $M=2$, $N=1$, and the particle initially localized on the left site.

The analytical solution of the Lindblad master equation follows from the vectorization procedure described in Appendix~\ref{app:linbland}. For the full Liouvillian evolution (including QJs), the expectation value of the site occupation $\langle n_1(t)\rangle$ is
\begin{equation}
\label{exact_n1}
\begin{split}
\langle n_1(t) \rangle =
  \frac{ e^{- \gamma t / 2} }{ \omega^2 }
  \Big[
    &\left( \gamma^2 - 8J^2 \right)
      \cosh\!\left( \frac{\omega t}{2} \right)
     - \gamma \omega
      \sinh\!\left( \frac{\omega t}{2} \right) \\
    &\quad - 8J^2
  \Big],
\end{split}
\end{equation}
where
\[
\omega = \sqrt{\gamma^2 - 16J^2}.
\]
The EP occurs at $\gamma = 4J$, where the above expression must be evaluated by taking the appropriate limit.

If QJs are omitted, the population dynamics becomes
\begin{equation}
\label{exact_n1_noQJs}
\langle n_1(t)\rangle =
1 -
\frac{16J^2 \sinh^2\!\left(\frac{\omega t}{4}\right)}
{\gamma^2 \cosh\!\left(\frac{\omega t}{2}\right)
- \gamma \omega \sinh\!\left(\frac{\omega t}{2}\right)
- 16J^2}.
\end{equation}

The LTA of the site occupation in the no-QJ case is
\begin{equation}
\label{exact_lta_noQJs}
\langle n_1(\gamma)\rangle_T =
\begin{cases}
\dfrac{1}{2}, & 0 < \gamma < 4J, \\
1 - \dfrac{8J^{2}}{\gamma^{2} - \gamma\sqrt{\gamma^{2} - 16J^{2}}}, & \gamma \ge 4J.
\end{cases}
\end{equation}
Perturbing near the EP, $\gamma = 4J + \delta$, with $\delta \ll 1$,
yields the square-root response
\begin{equation}
\label{exact_lta_noQJs2}
\langle n_1(\gamma)\rangle_T \approx
\dfrac{1}{2} - \frac{1}{2\sqrt{2J}}\sqrt{\delta},
\end{equation}
which directly reflects the EP2 nature of the degeneracy.

We now turn to the population imbalance defined in Eq.~\eqref{imbalance_definition}. Remarkably, for the noninteracting case this quantity is identical with or without QJs, since normalization factors cancel in the ratio. The exact time-dependent expression is
\begin{equation}
\label{exact_imbalance}
\langle z(t)\rangle =
1 -
\frac{32J^{2}\sinh^{2}\!\left(\dfrac{\omega t}{4}\right)}
{\gamma^{2}\cosh\!\left(\dfrac{\omega t}{2}\right)
- \gamma\omega\sinh\!\left(\dfrac{\omega t}{2}\right)
- 16J^{2}}.
\end{equation}

The corresponding LTA is
\begin{equation}
\label{lta_imbalance}
\langle z(\gamma)\rangle_T =
\begin{cases}
0, & 0 < \gamma < 4J, \\
-\dfrac{\sqrt{\gamma^{2} - 16J^{2}}}{\gamma}, & \gamma \ge 4J.
\end{cases}
\end{equation}

If we expand it similarly to what we did with the population, we find
\begin{equation}
\label{lta_imbalance_response}
\langle z(\gamma) \rangle_T \approx
-\frac{1}{\sqrt{2J}}\sqrt{\delta},
\end{equation}
demonstrating that the EP order can be inferred directly from imbalance dynamics.

\section{Dynamics before and after the EPs}
\label{app:Dyn}
\begin{figure}
\includegraphics[width=1\linewidth]{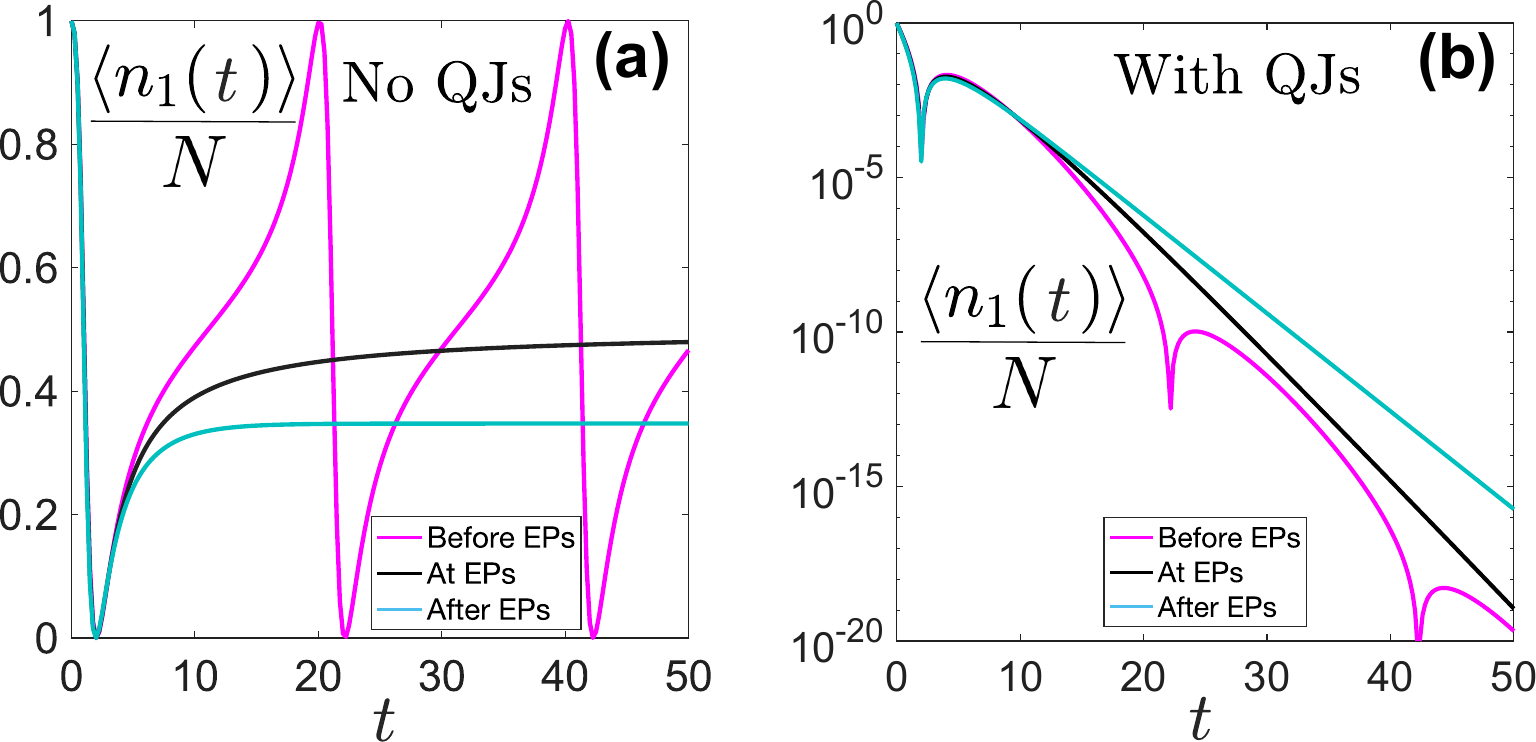}
\caption{Time evolution of $\langle n_1(t)\rangle$ before, at, and after the EP
($\gamma=2$) for the noninteracting Bose-Hubbard model.
(a) Dynamics without QJs.
(b) Full Liouvillian dynamics including QJs.}
\label{fig:Dynamics_BH}
\end{figure}

\begin{figure}
\includegraphics[width=1\linewidth]{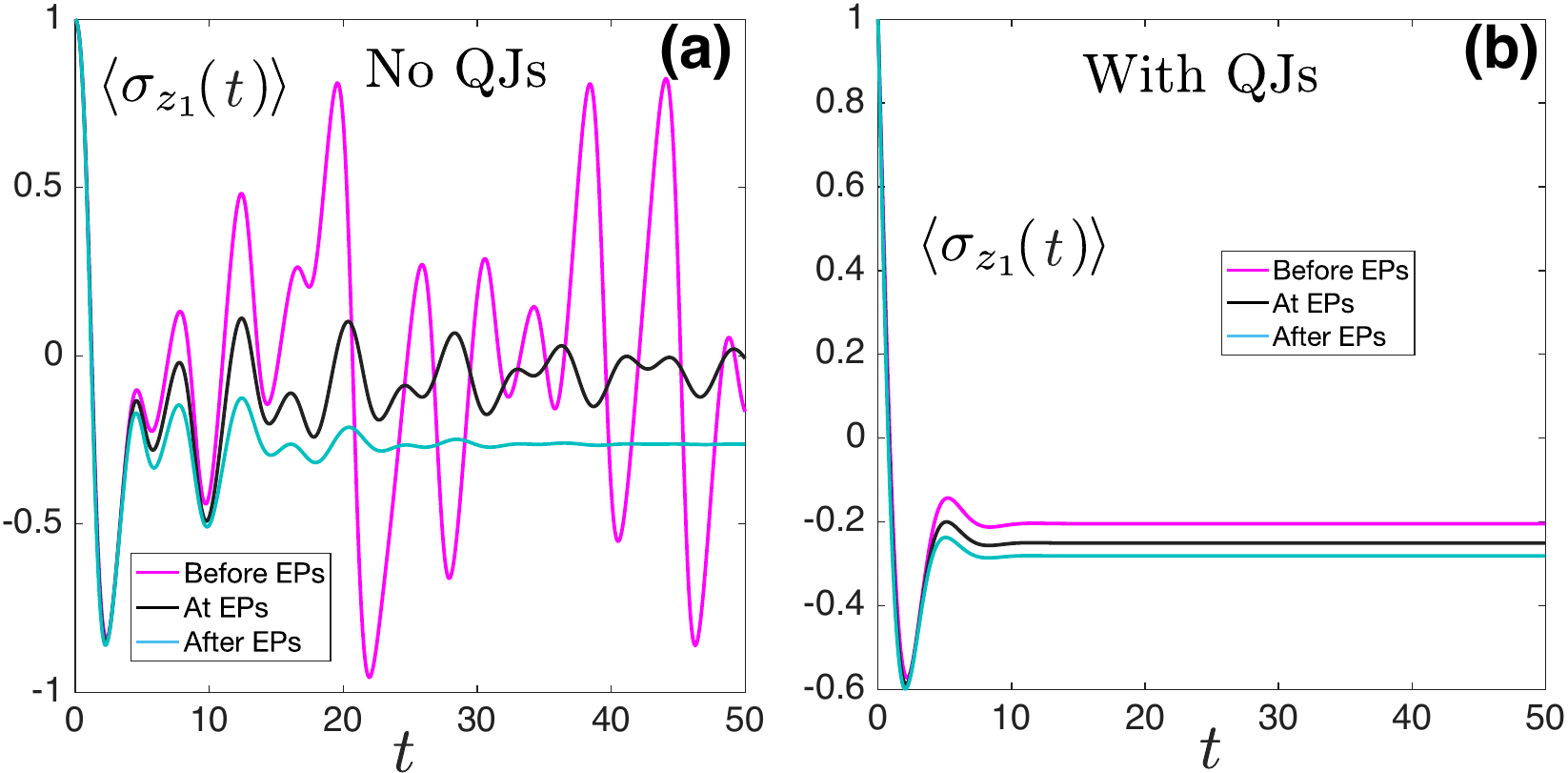}
\caption{Time evolution of $\langle \sigma_{z_1} (t) \rangle$ before, at, and after the EP for the two-spin model.
(a) Dynamics without QJs (EP at $\gamma \approx 1.076$).
(b) Full Liouvillian dynamics including QJs (EP at $\gamma \approx 0.689$).}
\label{fig:Dynamics_SC}
\end{figure}

This Appendix illustrates representative time evolutions before, at, and after EPs for both models. For the Bose-Hubbard dimer, see Fig. \ref{fig:Dynamics_BH} and for the spin chain, see Fig. \ref{fig:Dynamics_SC}. When QJs are omitted, coherent oscillations are present below the EP and are abruptly suppressed at the EP. Including QJs leads to dissipative dynamics across all dissipation strengths.

\section{Larger systems}
\label{app:large}

\begin{figure}
\includegraphics[width=1.0\linewidth]{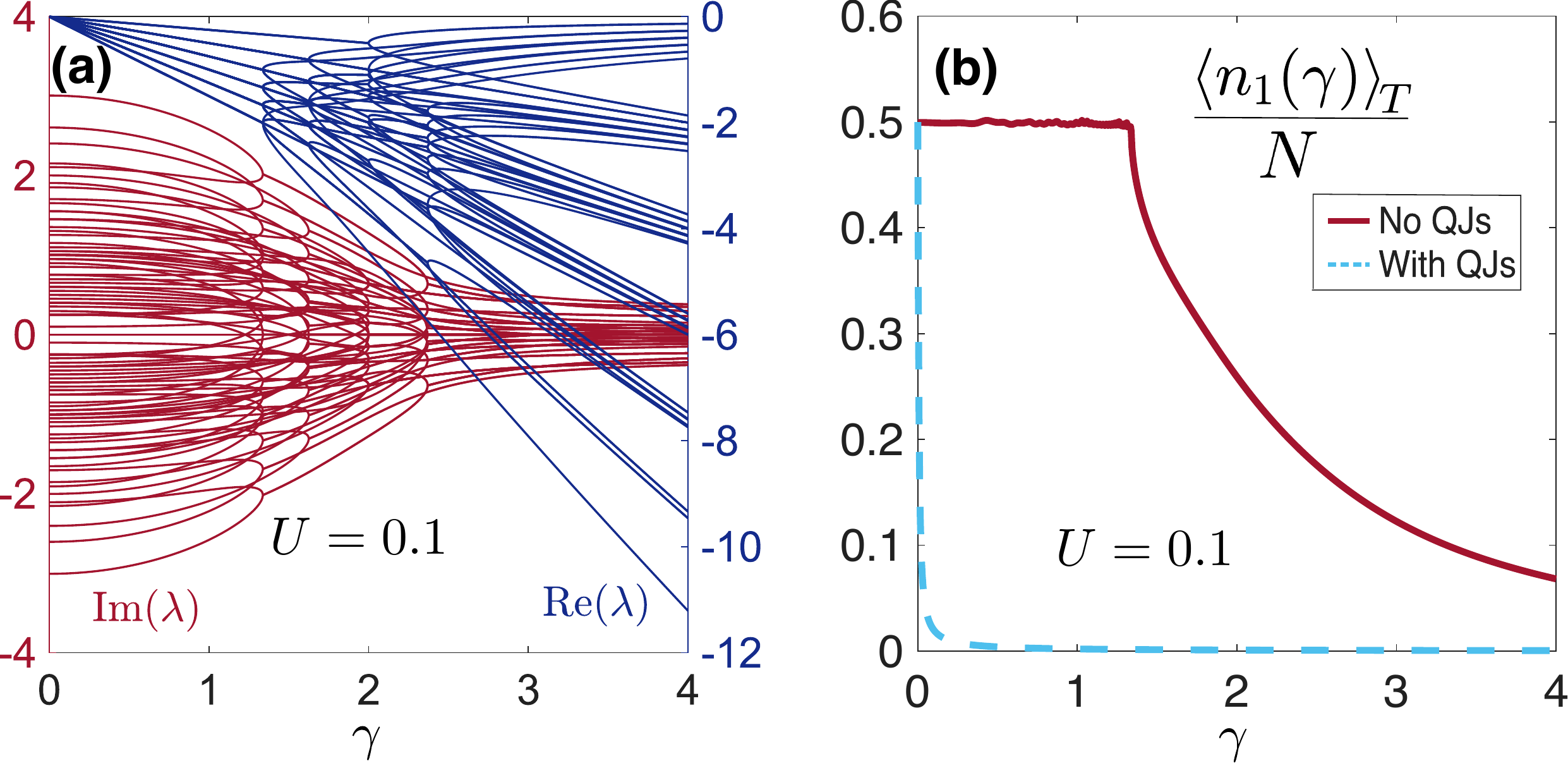}
\caption{Three bosons in the interacting Bose-Hubbard dimer.
(a) Liouvillian spectrum (identical with and without QJs).
(b) LTA $\langle n_1(\gamma)\rangle_T$. A cusp occurs in the no-QJ case at $\gamma \approx 1.338$. The population average is normalized by the initial particle number.}
\label{fig:3bosons}
\end{figure}

\begin{figure}
\includegraphics[width=1.0\linewidth]{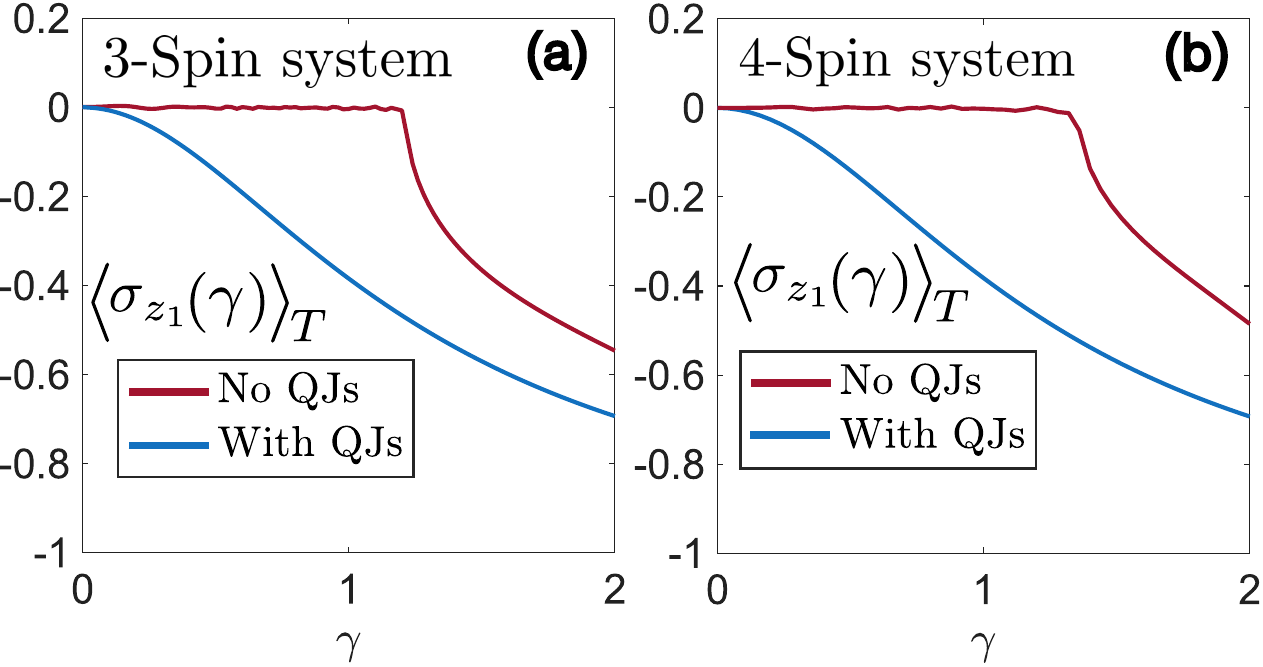}
\caption{LTA for larger spin chains.
(a) Three-spin chain.
(b) Four-spin chain.
Parameters: $\Omega=1$, $J=0.25$ with nearest-neighbor interactions. Cusps in the no-QJs case at $\gamma \approx 1.215$ for the three-spin and $\gamma \approx 1.356$ for the four-spin case, respectively.
}
\label{fig:spins34}
\end{figure}

In this Appendix, we present selected results for larger systems to demonstrate that the qualitative conclusions of the main text persist beyond minimal models. For the case of three bosons, see Fig. \ref{fig:3bosons}. { For larger spin chains, Fig.~\ref{fig:spins34} shows the corresponding LTA for three- and four-spin chains and supports the same qualitative no-QJ versus full-Lindblad distinction discussed in the main text with interaction only between first neighbors.}

\section{The quantum-jump method}
\label{app:qjm}


The quantum-jump method \cite{Breuer2007} is a numerical method that allows one an exact solution of Eq. \eqref{master_equation}. The advantage of this method is that instead of evolving the initial density matrix, $\hat{\rho}(0)=|\psi(0)\rangle\langle \psi(0)|$, we evolve the state vector along stochastic trajectories and average over their realizations. If we know the wavefunction at time $t$, then we propagate it to time $t+\delta t$ using the following stochastic procedure. We define the quantity ($\hbar=1$)
\begin{equation}
 \delta p = \sum_k \delta p_k,
\end{equation}
where
\begin{equation}
 \delta p_k = \langle \psi(t)|{L}_{k}^\dag {L}_k|\psi(t) \rangle \delta t
\end{equation}
is the probability of the process that is described by the Lindblad operator ${L}_k$ to occur. Consider also a uniformly distributed random number $\epsilon \in [0,1]$. 

If $\epsilon > \delta p$, the state $|\psi(t)\rangle$ is evolved continuously with the NHH defined in Eq. \eqref{nhh}:
\begin{equation}
 |\psi(t+\delta t)\rangle = \frac{1}{\sqrt{1-\delta p}}\left( 1 - i{H}_{\text{eff}}\delta t\right)|\psi(t)\rangle.
\end{equation}
If $\epsilon \leq \delta p$, the state performs a quantum jump and we choose the specific jump operator, ${L}_k$, according to the probability law $P_k = \delta p_k / \delta p$,
\begin{equation}
 |\psi(t+\delta t)\rangle = \frac{{L}_k |\psi(t)\rangle}{\sqrt{\delta p_k / \delta t}}.
\end{equation}

The number of Monte Carlo trajectories, $n$, should be large enough such that the averaged density matrix converges. The calculation of an observable $O$ is then trivial. One takes the mean value of the quantum average:
\begin{equation}
{\langle O (t)\rangle } = \frac{1}{n}\sum\limits_{i = 1}^n {\langle {\psi ^{(i)}}(t)|O|{\psi ^{(i)}}(t)\rangle }
\end{equation}
where $|{\psi ^{(i)}}(t)\rangle$ is the realization of the $i-$th trajectory.

\begin{figure}
\includegraphics[width=0.5\linewidth]{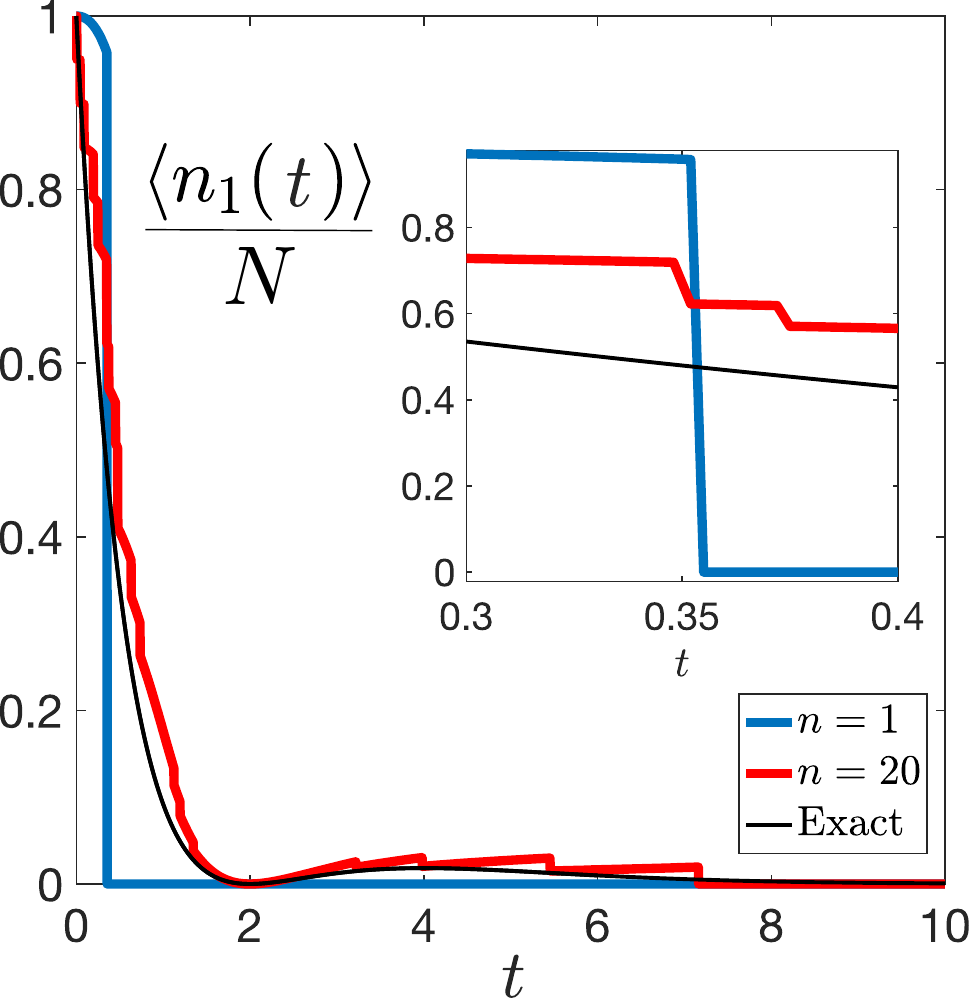}
\caption{Quantum-jump trajectories and averaged dynamics for the noninteracting Bose-Hubbard model.
Time evolution of the site occupation $\langle n_1(t) \rangle$ computed using the QJM.
Blue curve: A single stochastic trajectory, exhibiting a sharp quantum jump that removes the particle from the system.
Red curve: Average over $20$ quantum-jump trajectories, yielding a smooth evolution that approaches the exact result.
Black curve: Exact Liouvillian evolution obtained from Eq.~\eqref{exact_n1}.
Inset: Magnified view of a quantum-jump event corresponding to the action of the Lindblad operator
$L_1 = \sqrt{\gamma}\, a_1$, which describes particle loss from the left site.
}
\label{fig:qj_BH}
\end{figure}

Figure~\ref{fig:qj_BH} illustrates the origin of the effect of the QJs in the dynamics. A single stochastic trajectory (blue curve) exhibits a sudden quantum jump that removes the particle from the system. Averaging over multiple trajectories (red curve) smooths the evolution and reproduces the exact Liouvillian result (black curve). The inset highlights a quantum jump corresponding to the action of the Lindblad operator
$L_1 = \sqrt{\gamma} a_1$, describing particle loss from the left site. These results demonstrate that individual trajectories are highly discontinuous (demonstrating the stochastic nature of dissipation), while ensemble-averaged dynamics recover smooth dissipative behavior.

\begin{figure}
\includegraphics[width=1\linewidth]{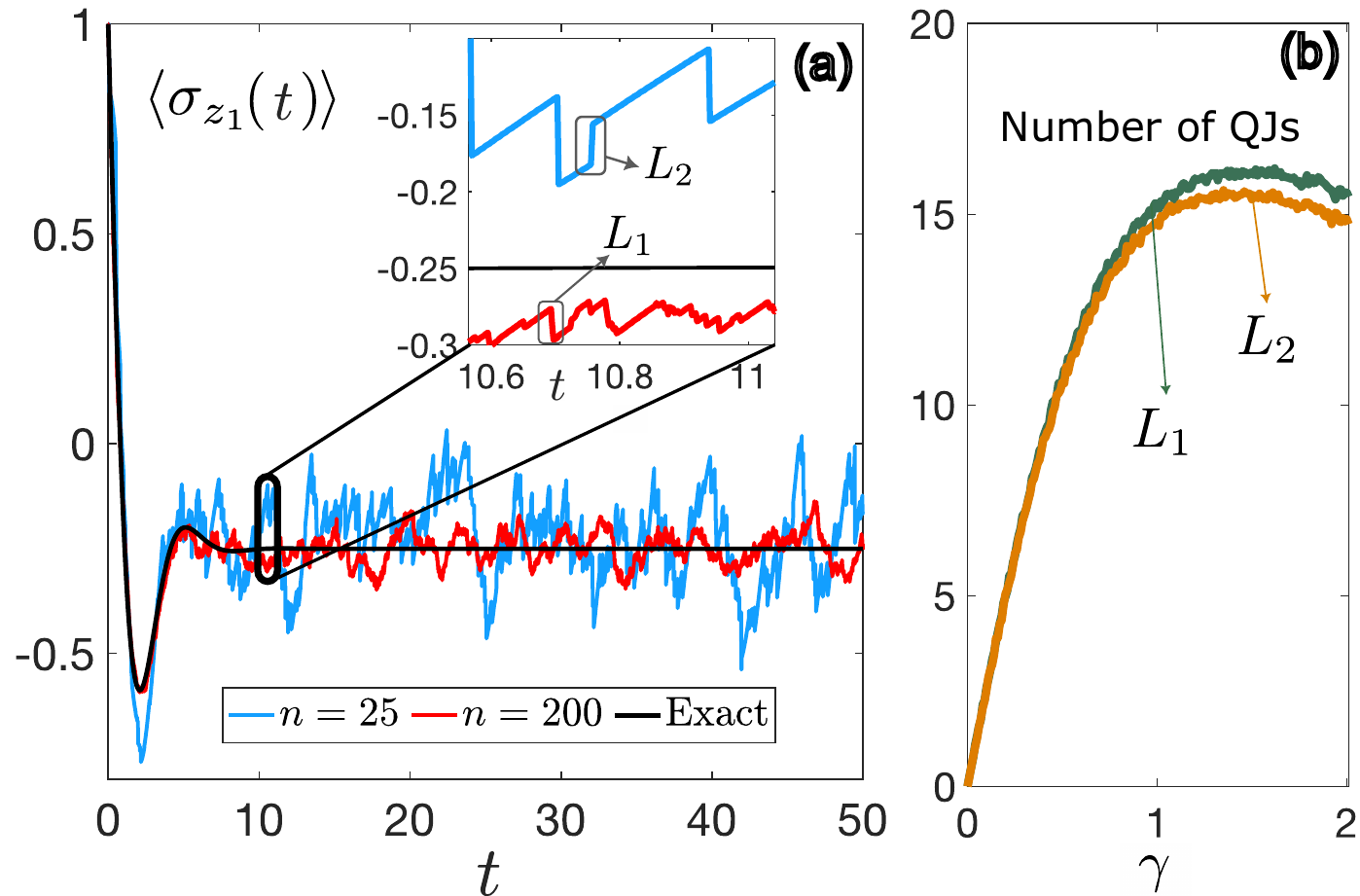}
\caption{Quantum-jump method for the two-spin system.
(a) Time evolution of $\langle \sigma_{z_1}(t)\rangle$ at the EP ($\gamma \approx 0.689$),
computed using the quantum-jump method with $n=25$ (blue) and $n=200$ (red) trajectories,
compared to the exact Liouvillian result (black). Increasing the number of trajectories suppresses stochastic fluctuations and improves agreement. Inset: Individual QJs associated with the Lindblad operators
$L_1$ (loss) and $L_2$ (gain).
(b) Total number of QJs due to $L_1$ and $L_2$ as a function of the dissipation rate $\gamma$, averaged over $1000$ trajectories. The number of QJs for $10$ and $100$ trajectories is shown in Fig. \ref{fig:Traj_both}.
}
\label{fig:qj_SC}
\end{figure}

\begin{figure}
\includegraphics[width=1\linewidth]{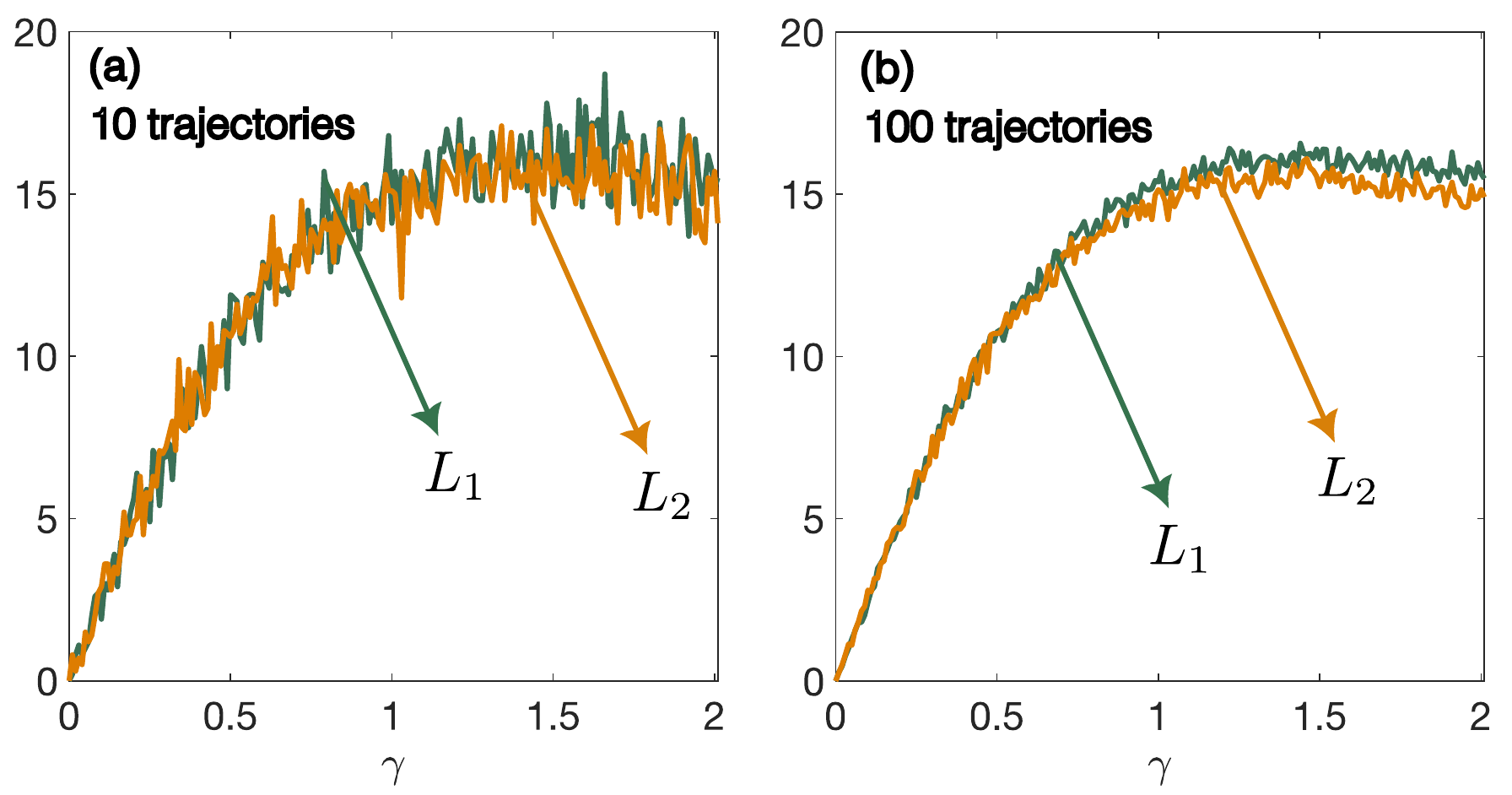}
\caption{Total number of QJs recorded for each process as a function of dissipation rate \( \gamma \) for $10$ trajectories [panel (a)] and $100$ trajectories [panel (b)]. The number of jumps increases with \( \gamma \), saturating at large values. The green curve corresponds to jumps due to \( L_1 \) (loss), while the orange curve corresponds to jumps due to \( L_2 \) (gain). }
\label{fig:Traj_both}
\end{figure}

Now, as mentioned in the main text, to further investigate the role of QJs in the dynamics of the two-spin system, we employ the QJM to simulate individual trajectories and compare them to the exact Liouvillian evolution. Figure~\ref{fig:qj_SC}(a) presents the time evolution of the observable \( \langle \sigma_{z_1} \rangle_t \) at the EP, computed using QJM with different numbers of trajectories. For \( n = 25 \) trajectories (blue), significant fluctuations are observed due to the stochastic nature of QJs. For \( n = 200 \) trajectories (red), the result is significantly smoother and closely matches the exact Liouvillian evolution (black). The inset provides a close-up view of individual QJs, showing discrete changes in \( \langle \sigma_{z_1} \rangle_t \) when a jump occurs due to either loss (\( L_1 \)) or gain (\( L_2 \)). These results demonstrate that increasing the number of trajectories improves convergence, reducing fluctuations and recovering the deterministic density matrix evolution. Figure~\ref{fig:qj_SC}(b) shows the total number of QJs per trajectory as a function of dissipation rate \( \gamma \). Both loss (\( L_1 \), green) and gain (\( L_2 \), orange) increase with \( \gamma \), reflecting stronger coupling to the environment. The number of jumps saturates at large \( \gamma \) (and begins to drop afterward), indicating that dissipation dominates the system’s dynamics.  Figure~\ref{fig:Traj_both} shows the quantum-jump counts obtained using fewer trajectories than in Fig.~\ref{fig:qj_SC}(b).

\label{app5}

\end{document}